\documentclass[12pt]{iopart}

\usepackage{graphicx}
\usepackage{bm} 
\usepackage{subfigure}
\usepackage[colorlinks=true,linktocpage=true,linkcolor=blue,citecolor=blue]{hyperref}
\usepackage[usenames,dvipsnames]{color}
\usepackage[square,sort,comma,numbers,sort&compress]{natbib}

\usepackage{amssymb}

\def\pt{p_\perp}
\def\pl{p_\parallel}

\begin{document}
%



\title{Thermalization of parton spectra in the color-flux-tube model}


\author{Radoslaw Ryblewski}
 \ead{Radoslaw.Ryblewski@ifj.edu.pl}

\address{The H. Niewodnicza\'{n}ski Institute of Nuclear Physics, Polish Academy of Sciences, PL-31-342 Krak\'{o}w, Poland}

\begin{abstract}
Detailed study of thermalization of the momentum spectra of partons produced via decays of  the color flux tubes due to the Schwinger tunneling mechanism is presented. The collisions between particles are included in the relaxation time approximation specified by different values of the shear viscosity to entropy density ratio. At first we show that, to a good approximation, the transverse-momentum spectra of the produced patrons are exponential, irrespectively from the  assumed value of the viscosity of the system and the freeze-out time. This thermal-like behaviour may be attributed to specific properties of the Schwinger tunneling process. In the next step, in order to check the approach of the system towards genuine local equilibrium, we compare the local slope of the model transverse-momentum spectra with the local slope of the fully equilibrated reference spectra characterised by the effective temperature that reproduces the energy density of the system. We find that the viscosity corresponding to the AdS/CFT lower bound is necessary for thermalization of the system within about two fermis. 
\end{abstract}

\date{2 March 2015}

%
%
%
%
%


\section{Introduction}
\label{sec-intro}
%
\par The properties and space-time evolution of strongly-interacting matter produced in ultra-relativistic heavy-ion collisions at the Relativistic Heavy-Ion Collider (RHIC) and the Large Hadron Collider (LHC) are subject to intense experimental and theoretical studies for more than decade now. The detailed analyses of hadronic observables suggest that the initial processes in a short time of about one fermi give rise to a relatively well equilibrated system of elementary particles, that has been  named the Quark-Gluon Plasma (QGP) \cite{Shuryak:1978ij}. Due to its non-Abelian structure,  the underlying theory of such processes, Quantum Chromodynamics (QCD), is far too complex to study  the system's real time dynamics. Instead, it is argued that the system, if thermalized sufficiently early, may be subsequently described in  the  classical framework of fluid dynamics \cite{Muronga:2003ta,Heinz:2005bw,Bhalerao:2005mm,Baier:2006um,Baier:2007ix,
Dusling:2007gi,Broniowski:2008vp,NoronhaHostler:2008ju,
Bozek:2009dw,El:2009vj,Alver:2010dn,Martinez:2010sc,Florkowski:2010cf,
PeraltaRamos:2010je,Petersen:2010cw,
Denicol:2010tr,Pratt:2010jt,
Schenke:2010rr,
Niemi:2011ix,Shen:2011eg,Ryblewski:2012rr,Pang:2012he,
Karpenko:2015xea,Bhalerao:2015iya,Becattini:2015ska,Bazow:2015cha,Tinti:2015xwa}, 
with initial conditions determined within some microscopic model of the initial stage, such as Color Glass Condensate (CGC)/Glasma \cite{McLerran:1993ni,Iancu:2003xm,JalilianMarian:2005jf,Gelis:2009wh}, Monte-Carlo Glauber model \cite{Rybczynski:2013yba} or IP-Glasma \cite{Schenke:2012wb}. 

The fast thermalization, required for subsequent application of viscous fluid dynamics in such a system is, however, hard to prove and, in fact, it has been a subject of intensive studies over the last years, that identified in particular an important role of plasma instabilities \cite{Mrowczynski:1993qm,Randrup:2003cw,Romatschke:2003ms,Arnold:2003rq,Mrowczynski:2004kv,Rebhan:2008uj,Ipp:2010uy}. Nevertheless, a significant progress has been done recently in this subject using microscopic approaches, both in the weak \cite{Gelis:2013rba,Berges:2013eia} and strong  \cite{Chesler:2010bi,CaronHuot:2011dr,Heller:2011ju,vanderSchee:2013pia} coupling limits. Qualitatively, similar results have been also found recently in Ref.~\cite{Ryblewski:2013eja} in a simpler framework of the color-flux-tubes model \cite{Casher:1978wy,Glendenning:1983qq,Bialas:1984wv,Bialas:1984ap,
Bialas:1985is,Gyulassy:1986jq,Gatoff:1987uf}, where the color field dynamics is treated in the Abelian dominance approximation \cite{Heinz:1983nx,Heinz:1984yq,Elze:1986qd,Elze:1986hq,Bialas:1986mt,Bialas:1987en,Dyrek:1988eb} (these results have been reproduced in Ref.~\cite{Ruggieri:2015yea}  in the Monte-Carlo transport approach \cite{Plumari:2012ep,Scardina:2014gxa,Scardina:2014nja,Puglisi:2014sha,Das:2015aga,Ruggieri:2015tsa,Plumari:2015cfa}). In Ref.~\cite{Ryblewski:2013eja} the production of partons appears in the model through the decay of color fields by the Schwinger tunneling  mechanism \cite{Schwinger:1951nm,Brezin:1970xf,Casher:1978wy,Glendenning:1983qq,
Gyulassy:1986jq,Schmidt:1998vi,Fukushima:2009er,Gelis:2013oca,Gelis:2015kya} and the thermalization is included explicitly through the introduction of the collisional kernels treated in the relaxation-time approximation (RTA) \cite{Bhatnagar:1954zz,Baym:1984np,Baym:1985tna,Banerjee:1989by,
Heiselberg:1995sh,Wong:1996va,Nayak:1996ex,Nayak:1997kp,Bhalerao:1999hj}.
\par In this paper we present a detailed study of the early-time thermalization of the spectrum of the plasma constituents in relativistic heavy-ion collisions using the framework described in Ref.~\cite{Ryblewski:2013eja}. At first, we find that without the collisions the system, although showing  thermalized (i.e., of the exponential shape) transverse-momentum, $\pt$, spectra, is free-streaming, with the proper-time dependence of the effective temperature described very well by the exact (0+1)-dimensional [(0+1)D] free-streaming solution found within anisotropic hydrodynamics in Refs.~\cite{Martinez:2010sc,Florkowski:2010cf}. Hence, this case may be described as {\it apparent two-dimensional thermalization}. In the next step, we argue that the inclusion of collisions with the shear viscosity to entropy density ratio, $\bar{\eta}=\eta/s$, corresponding to the AdS/CFT lower bound of $1/(4\pi)$  \cite{Policastro:2001yc,Kovtun:2004de}  is necessary to fully thermalize the system within about two fermis, which is required for subsequent application of viscous fluid dynamics. By {\it full thermalization} in this context we mean reaching local equilibrium, where the system is isotropic in the momentum space and the three-dimensional spectra correspond to equilibrium distributions. Finally, we show that the deviations of the exact spectra predicted by our model from the fully thermalized ones may be very well handled with the usage of the standard viscous hydrodynamics  phase-space distribution function obtained within Grad 14-moment approximation. We identify this fact with the so called \textit{hydrodynamization} of the system \cite{Heller:2011ju}.

\par The  structure  of  the  paper  is  as  follows:  In  Sec.~\ref{sect-sym}  we  introduce the Bjorken symmetry and several parametrizations used in the calculations.  In Sec.~\ref{sect-kin} we review the version of the color-flux-tube model introduced first in Ref.~\cite{Ryblewski:2013eja}.  In Sec.~\ref{sect-spec} we derive the formulas for the transverse-momentum spectra of partons using the Cooper-Frye formula. In Sec.~\ref{sect-res} we discuss our results for the collisionless plasma and plasma with collisions included. We summarize in Sec.~\ref{sect-con}. In the paper we use natural units where $c= 1$, $k_B= 1$, and $\hbar= 1$.

\section{Imposing Bjorken symmetry}
\label{sect-sym}
%
In the case of (0+1)D, longitudinally boost-invariant and transversally homogeneous expansion (commonly referred to as the Bjorken model~\cite{Bjorken:1982qr}), it is convenient to introduce Milne parameterization of Minkowski space-time coordinates
\begin{eqnarray}
x^\mu &=& \left(t, {\vec x}_\perp, z \right) =
\left(\tau \cosh \eta, {\vec x}_\perp, \tau \sinh \eta \right) ,
\label{parx}
\end{eqnarray} 
where $\tau \equiv \sqrt{t^2 - z^2}$ is the longitudinal proper time and $\eta \equiv \tanh^{-1}\left(z/t\right)$ is the space-time rapidity. The corresponding boost-invariant  parameterization of the particle four-momentum is
\begin{eqnarray}
p^\mu &=& \left(E, {\vec p}_\perp, p_\parallel \right) =
\left(m_\perp \cosh y,  {\vec p}_\perp, m_\perp \sinh y \right),
\label{parp}
\end{eqnarray} 
where \mbox{$m_\perp \equiv \sqrt{m^2 + \pt^2}$} is the transverse mass and \mbox{$y \equiv \tanh^{-1}\left(E/p_\parallel\right)$} is the rapidity. For the particles on the mass shell we have $p^2 =E^2-{\vec p}^{\,\,2}=m^2$, where $m$ is the particle rest mass. The flow of matter is fixed by the Bjorken symmetry and has the form \cite{Bjorken:1982qr}
\begin{equation}
u^\mu = \left(t/\tau,0,0,z/\tau\right).  \label{u}
\end{equation}
We introduce also the other two convenient boost-invariant variables \cite{Bialas:1984wv} which mix space-time and four-momentum coordinates, namely
\begin{eqnarray}
w &=&  t p_\parallel - z E = \tau m_\perp \sinh \left(y- \eta\right),\\
v &=& Et-p_\parallel z = \tau m_\perp \cosh \left(y- \eta\right).
\label{wv}
\end{eqnarray}
We note that  $v^2 - w^2 = m_\perp^2 \tau^2$. 

The requirement of boost-invariance implies that all scalar functions of space and time depend on $\tau$ solely, whereas the phase-space distribution function may depend only on  $\tau$, $w$ and $p_\perp$ \cite{Bialas:1984wv}, i.e., $f(x, p) = f(\tau,w,p_\perp)$.  The integration measure in the momentum sector of the phase-space is \mbox{$
dP = dp_{\Vert }d^2p_{\bot }/p^0=dw\, d^2p_{\bot } /v$}.
In what follows we neglect masses of quarks and set $m=0$. 
%
\section{Quark-gluon plasma dynamics in the color-flux-tube approach}
\label{sect-kin}
%
Henceforth, we consider a quark-gluon plasma whose dynamics in the Abelian dominance approximation \cite{Heinz:1983nx,Heinz:1984yq,Elze:1986qd,Elze:1986hq,Bialas:1986mt,Bialas:1987en,Dyrek:1988eb} may be described within the following transport equations 
\begin{equation}
\left( p^\mu \partial_\mu + g{\mbox{\boldmath $\epsilon$}}_i \cdot 
{\bf F}_{}^{\mu \nu } p_\nu \partial_\mu^p\right) f_{if}(x,p) = \frac{dN_{if}}{d\Gamma_{\rm inv} }+ C_{if},  
\label{kineq}
\end{equation}
\begin{equation}
\left( p^\mu \partial_\mu - g{\mbox{\boldmath $\epsilon$}}_i \cdot 
{\bf F}_{}^{\mu \nu } p_\nu \partial_\mu^p\right) \bar{f}_{if}(x,p) = \frac{dN_{if}}{d\Gamma_{\rm inv} } + \bar{C}_{if},  
\label{kineaq}
\end{equation}
\begin{equation}
\hspace{-0.12cm}\left( p^\mu \partial _\mu + g{\mbox{\boldmath $\eta$}}_{ij} \cdot 
{\bf F}_{}^{\mu \nu } p_\nu \partial_\mu^p\right) \widetilde{{f}}_{ij}(x,p) = 
\frac{d\widetilde{N}_{ij}}{d\Gamma_{\rm inv} }+ \tilde{C}_{ij}, \label{kineg}
\end{equation}
for quark, antiquark, and charged-gluon single-particle phase-space distribution functions, respectively, see Ref.~\cite{Ryblewski:2013eja}. The terms on the left-hand sides of Eqs.~(\ref{kineq})--(\ref{kineg}) describe the free-streaming of particles and their interaction with the mean color field, ${\bf F}^{\mu\nu}=\left(F^{\mu\nu}_{(3)},F^{\mu\nu}_{(8)} \right)$, with the only non-vanishing component corresponding to the longitudinal chromoelectric field, ${\bf F}^{30} ={\cal E}$. The partons couple to the field  through the charges $\mbox{\boldmath $\epsilon$}_{i}$ (for quarks), $-\mbox{\boldmath $\epsilon$}_{i}$ (for antiquarks), and ${\mbox{\boldmath $\eta$}}_{ij}={\mbox{\boldmath $\epsilon$}}_{i}-{\mbox{\boldmath $\epsilon$}}_{j}$ (for gluons) \cite{Huang:1982ik}, where the colour indices $i,j$ run from 1 to 3.

The first terms on the right-hand sides of Eqs.~(\ref{kineq})--(\ref{kineg}) describe the particle production due to the decay of the color field through the Schwinger tunneling  mechanism \cite{Schwinger:1951nm,Brezin:1970xf,Casher:1978wy,Glendenning:1983qq,Gyulassy:1986jq} and they are given by the formula $dN/d\Gamma_{\rm inv} =  {\cal R} \delta(p_\parallel) p^0$, where 
\begin{eqnarray}
 \label{rateq}
 {\cal R}_{if}&=&\frac{\Lambda _i}{4\pi ^3}\left| \ln \left( 1-\exp
\left( -\frac{\pi m_{f\perp }^2}{\Lambda _i}\right) \right) \right|, \\
 \widetilde{{\cal R}}_{ij}&=&\frac{ \widetilde{\Lambda} _{ij}}{4\pi ^3}\left| \ln
\left( 1+\exp \left( -\frac{\pi p_{\bot }^2}{ \widetilde{\Lambda} _{ij}}\right) \right)
\right|,
 \label{rateg}
\end{eqnarray}
and
\begin{eqnarray}
\Lambda _i&=&\left( g\left| {\mbox{\boldmath $\epsilon$}}_i\cdot {%
\mbox{\boldmath $\cal E$}}\right| -\sigma _q\right) \theta \left( g\left| {%
\mbox{\boldmath $\epsilon$}}_i\cdot {\mbox{\boldmath $\cal E$}}\right|
-\sigma _q\right),\\
 \widetilde{\Lambda} _{ij}&=&\left( g\left| {\mbox{\boldmath $\eta$}}_{ij}\cdot {%
\mbox{\boldmath $\cal E$}}\right| -\sigma _g\right) \theta \left( g\left| {%
\mbox{\boldmath $\eta$}}_{ij}\cdot {\mbox{\boldmath $\cal E$}}\right|
-\sigma _g\right),   \label{Lami}
\end{eqnarray}
where $\sigma_{q(g)}$ is the string tension for quarks (gluons), $g$ is the strong coupling constant, $\theta$ is the step function and index $f$ denotes the quark flavour. The last terms in Eqs.~(\ref{kineq})--(\ref{kineg}) describe the collision terms which herein are treated in the relaxation time approximation \cite{Bhatnagar:1954zz,Baym:1984np,Baym:1985tna,Heiselberg:1995sh,Wong:1996va}
\begin{eqnarray}
 C_{if} = p \cdot u \, \frac{f^{\rm eq}-f_{if}}{\tau_{\rm eq}}, \quad
\bar{C}_{if} = p \cdot u \, \frac{f^{\rm eq}-\bar{f}_{if}}{\tau_{\rm eq}},
\quad\widetilde{C}_{ij} = p \cdot u \, \frac{f^{\rm eq}-\tilde{f}_{ij}}{\tau_{\rm eq}},
\label{col-terms}
\end{eqnarray}
where 
\begin{eqnarray}
f^{\rm eq}(x,p) = \frac{g_s}{(2\pi)^3} 
\exp\left(-\frac{p \cdot u(x)}{T(x)} \right),
\label{feq}
\end{eqnarray}
with $g_s =2$ being the spin degeneracy. The relaxation time is expressed through the  shear viscosity  to entropy density ratio ${\bar \eta}$ as $\tau_{\rm eq}(\tau) = 5 {\bar \eta}/T(\tau)$  \cite{Anderson1974466,Czyz:1986mr,Dyrek:1986vv,
Romatschke:2011qp,Florkowski:2013lza,Florkowski:2013lya}.

It is straightforward to check that in the case of (0+1)D Bjorken expansion the kinetic equations (\ref{kineq})--(\ref{kineg}) have the following formal solutions \cite{Ryblewski:2013eja}
\begin{eqnarray}
\hspace{-1.8cm}f_{if}\left(\tau,w,p_{\bot }\right)\!\!\!&=&\!\!\! \int\limits_0^\tau d\tau^\prime D(\tau,\tau^\prime) \left[ \tau^\prime 
{\cal R}_{if}(\tau^\prime,p_\perp) 
\delta(\Delta h_i + w)+ \frac{f^{\rm eq}(\tau^\prime,\Delta h_i + w,p_\perp)}{\tau_{\rm eq}(\tau^\prime)}
\right], \nonumber \\
\hspace{-1.8cm}{\bar f}_{if}\left(\tau,w,p_{\bot }\right)\!\!\!&=&\!\!\! \int\limits_0^\tau d\tau^\prime D(\tau,\tau^\prime) \left[ \tau^\prime 
{\cal R}_{if}(\tau^\prime,p_\perp) 
\delta(\Delta h_i - w) 
+ \frac{{f}^{\rm eq}(\tau^\prime,\Delta h_i - w,p_\perp)}{\tau_{\rm eq}(\tau^\prime)}
\right], \label{formsol} \\
\hspace{-1.8cm}{\widetilde f}_{ij}\left(\tau,w,p_{\bot }\right)\!\!\!&=&\!\!\!\int\limits_0^\tau d\tau^\prime D(\tau,\tau^\prime) \left[ \tau^\prime 
{\widetilde{\cal R}}_{ij}(\tau^\prime,p_\perp) 
\delta(\Delta h_{ij} + w) 
+ \frac{f^{\rm eq}(\tau^\prime,\Delta h_{ij} 
+ w,p_\perp)}{\tau_{\rm eq}(\tau^\prime)}
\right], \nonumber 
\end{eqnarray}
where we introduced the damping function $D(\tau_2,\tau_1) = \exp\left(-\int\limits_{\tau_1}^{\tau_2}
d\tau^{\prime\prime}/\tau_{\rm eq}(\tau^{\prime\prime}) \right)$. The functions $\Delta h$, which are eventually integrated out when calculating the spectra of particles (see Eq.~(\ref{cfbi})), are defined in \cite{Ryblewski:2013eja}. 
\par The initial chromoelectric field spanned by the colliding nuclei follows from the Gauss law applied to single color flux tube and is set by  the condition ${\cal E} = \sqrt{2\sigma_{g}/{\cal A}} \,k\, {\bf q}$, where $\sigma_{g} = 3\,\sigma_{q} = 3\,{\rm  GeV/fm}$, ${\cal A} = \pi R_\perp^2 = 1\,{\rm fm^2}$ is the transverse area of a tube, $k=10$ \cite{Ryblewski:2013eja} is the number of color charges spanning the tube, and ${\bf q}$ is the color charge of quark or gluon. In consequence the strong coupling constant is $g = \sqrt{6\,{\cal A}\,{\rm GeV/fm}} \approx 5.48$.
%
\section{Transverse-momentum spectra}
\label{sect-spec}
%
The transverse-momentum spectra may be calculated from the Cooper--Frye
formula \cite{Cooper:1974mv}
\begin{equation}
\frac{dN}{dy\, d^2p_\perp } = \int d\Sigma_\mu(x) p^\mu \,
f(x,p),
\label{cf}
\end{equation}
where $d\Sigma_\mu(x)$ is the element of the
freeze-out hypersurface which may be obtained with the help of the expression known from differential geometry \cite{Misner:1974qy}
\begin{equation}
d\Sigma_\mu = \varepsilon_{\mu \alpha \beta \gamma}
\frac{\partial x^\alpha}{\partial x} \frac{\partial x^\beta}{\partial y} \frac{\partial x^\gamma}{\partial \eta }
d x \,d y \,d  \eta  \, .
\label{d3sigma}
\end{equation}
Here $\varepsilon_{\mu \alpha \beta \gamma}$ is the totally antisymmetric Levi--Civita tensor with $\varepsilon_{0123} = 1$. Assuming that the freeze-out occurs on the constant proper time hypersurface defined by the condition $\tau = {\rm const}$ we get $d\Sigma_\mu=u_\mu \tau\,dx \,dy \,d\eta $. Since the system is transversally homogeneous we may arbitrarily set the transverse radius, $R_\perp$, of the system in such a way that the integration (\ref{cf}) gives the production per unit area, \mbox{${\cal A} = 1~{\rm fm}^2$}. Using Eqs.~(\ref{parp}), (\ref{u}), and (\ref{wv}) we have $p \cdot u = v/\tau$, which allows us to rewrite Eq.~(\ref{cf}) in the form
\begin{eqnarray}
\frac{dN}{dy\, d^2p_\perp } =\pi R_\perp^2 \int\limits_{-\infty}^{+\infty} dw 
\,f(\tau,w,p_\perp),
\label{cfbi}
\end{eqnarray}
where we changed the integration measure using \mbox{$v d\eta = d w$}. In the special case, where the system is locally in the equilibrium state,  the distribution function has the form (\ref{feq}) leading to the following transverse-momentum spectrum 
\begin{eqnarray}
\frac{dN}{dy\, d^2p_\perp } =  \frac{g_s R_\perp^2}{(2 \pi)^2} \tau p_\perp K_1\left(\frac{p_\perp}{T(\tau)}\right),
\label{cfeq}
\end{eqnarray}
where $K_1$ is the modified Bessel function of the second kind. We note here that for \mbox{$\pt \gg T$} Eq.~(\ref{cfeq}) scales as 
$\sqrt{\pt T}\exp\left(-\pt/T\right)$, hence,  to a good approximation it is an exponential.

In what follows we assume that the system distribution function is the sum of the quark and gluon distribution functions (\ref{formsol}) 
\begin{eqnarray}
\hspace{-1.6cm}f\left(\tau,w,p_{\bot }\right) =  \sum_f^{N_f} \sum_i^3  \left( f_{if}\left(\tau,w,p_{\bot }\right) +   \bar{f}_{if}\left(\tau,w,p_{\bot }\right)\right) 
+\sum_{i\neq j=1}^3  \widetilde{f}_{ij}\left(\tau,w,p_{\bot }\right).
\label{f}
\end{eqnarray}
Using Eq.~(\ref{f}) in (\ref{cfbi}) we get
\begin{eqnarray}
\hspace{-0.25cm}\frac{dN}{dy\, d^2p_\perp }
&=&\frac{R_\perp^2}{2 \pi^2} \int\limits_0^\tau \tau^\prime d\tau^\prime D(\tau,\tau^\prime) \left\{ \frac{3 g_s (N_f +1)}{\tau_{\rm eq}(\tau^\prime)} \, p_\perp K_1\left(\frac{p_\perp}{T(\tau^\prime)}\right)\right.\nonumber\\ \nonumber
& & \hspace{-1cm}+ \left.   N_f \sum_{\rm i = 1}^3 \Lambda_i(\tau^\prime) \left| \ln \left( 1-\exp
\left( -\frac{\pi m_{f\perp }^2}{\Lambda _i(\tau^\prime)}\right) \right) \right|   \right. \\
& & \hspace{-1cm}\left. +\sum_{\rm i>j = 1}^3 
\,{\widetilde \Lambda} _{ij} (\tau^\prime)\left| \ln
\left( 1+\exp \left( -\frac{\pi p_{\bot }^2}{{\widetilde\Lambda} _{ij} (\tau^\prime)}\right) \right)
\right| \right\},
\label{cfmodel}
\end{eqnarray}
where in the numerical calculations we take $N_f =2$.
%
\section{Results}
\label{sect-res}
%

In this Section we analyse the momentum spectra of partons produced in the system defined above. In order to have the reference case, we first consider the plasma without collisions, which is formally obtained by setting the ratio $\eta/s$ to infinity. Subsequently, we move to the discussion of the influence of collisions on the system behaviour. The effect of collisions is included by the RTA collisional kernels.
%
\subsection{Collisionless plasma}
\label{sect-colles}
\begin{figure}[h]
\begin{center}
\includegraphics[width=0.49 \textwidth]{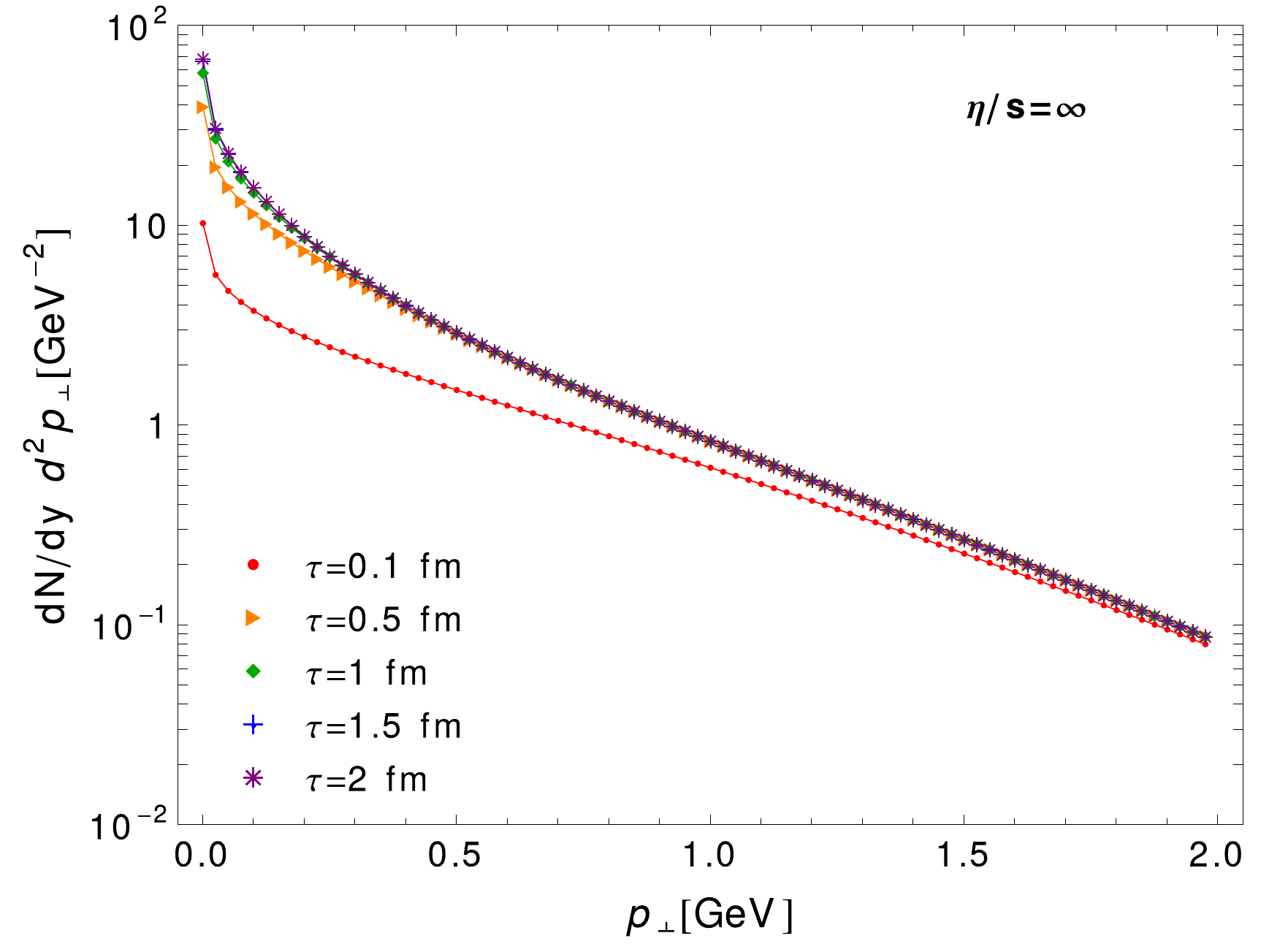}
\includegraphics[width=0.49 \textwidth]{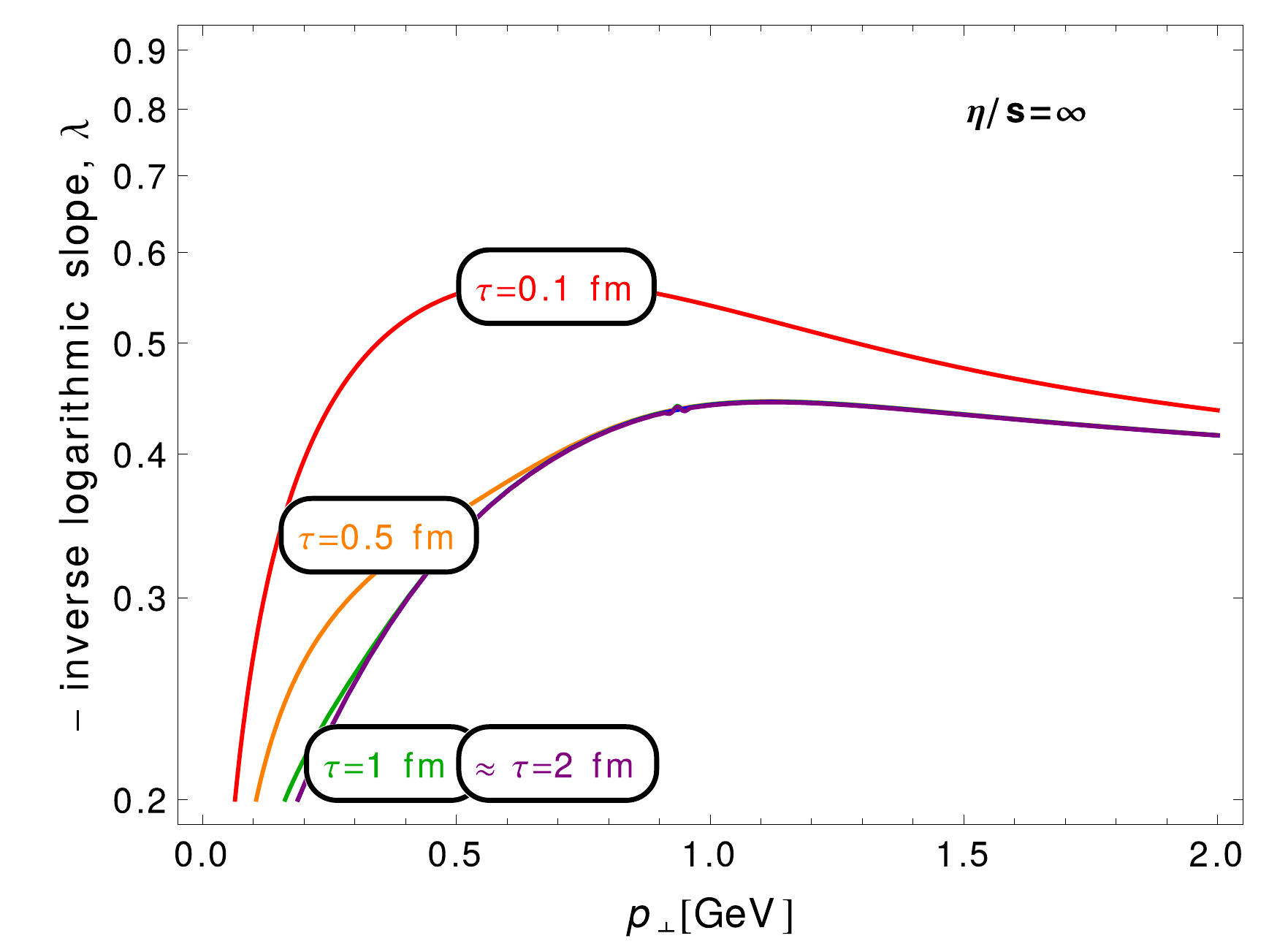}
\vspace{-0.2cm}
\end{center}
\caption{
The transverse-momentum spectra of partons obtained from Eq.~(\ref{cfmodel}) with the collisions excluded shown for various freeze-out proper times (left). The corresponding inverse slope parameter $\lambda$ (right), as introduced in Eq.~(\ref{slope}).
}
\label{sp0}
\end{figure}

%
Let us first consider the plasma production described by Eq.~(\ref{cfmodel}) excluding the collisions of particles, which corresponds to setting $\bar{\eta} = \infty$\footnote{In this case the first term in curly brackets may be dropped.}. In this case, already after a short time, \mbox{$\tau\approx 0.5-1$ fm}, the shape of the parton spectrum becomes frozen, see the left panel of Fig.~\ref{sp0}. In order to study the proper-time evolution of the spectra in more detail,  in the right panel of Fig.~\ref{sp0} we plot the \textit{negative inverse logarithmic slope} $\lambda$ of the spectrum defined in the following way
\begin{equation}
\lambda = -\left[\frac{d}{d\pt} \ln\left(\frac{dN}{dy\, d^2p_\perp }\right)\right]^{-1}.\label{slope}
\end{equation}
%
%
%
%
\begin{figure}[h]
\begin{center}
\includegraphics[width=0.49 \textwidth]{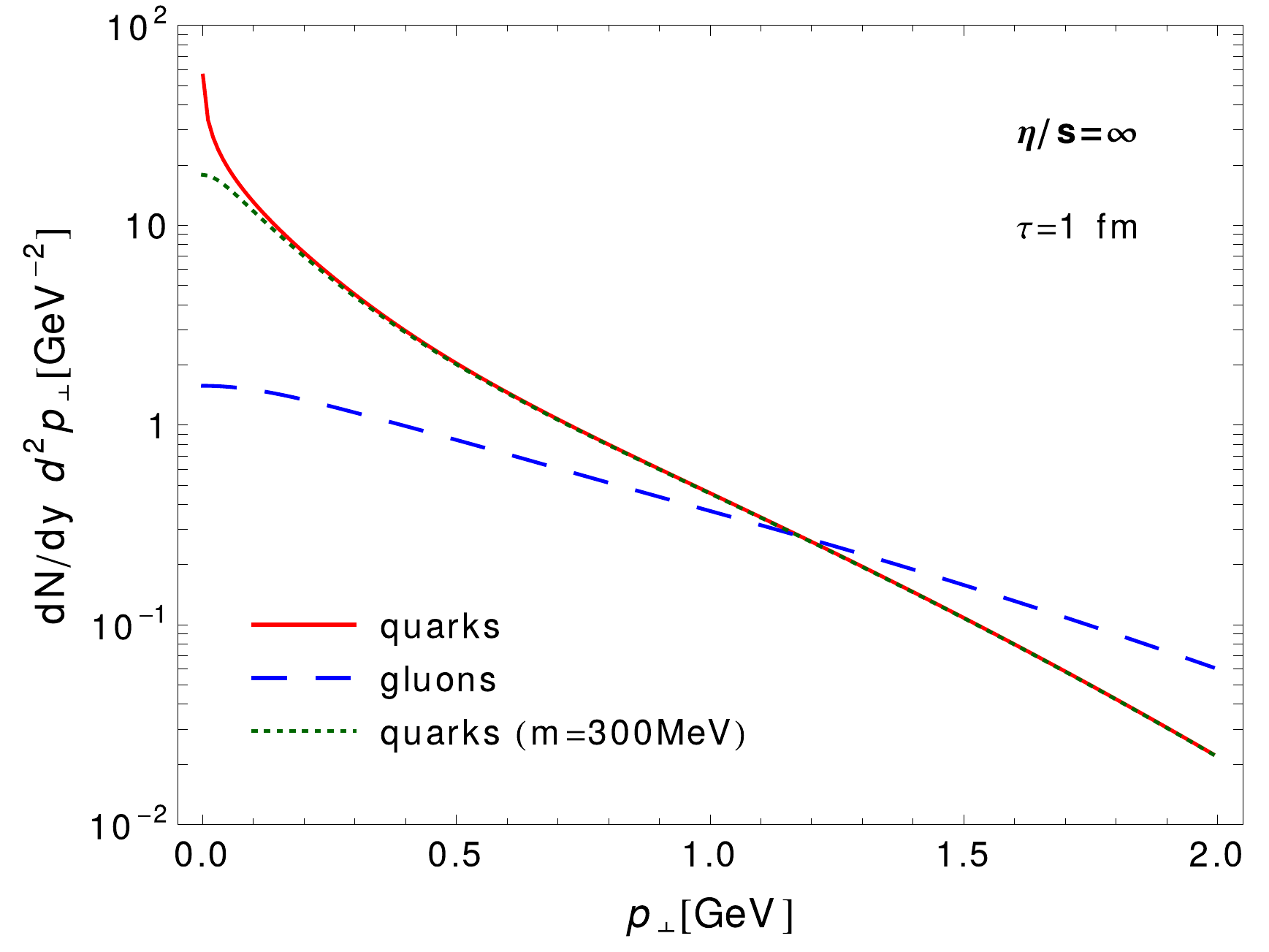}
\includegraphics[width=0.49 \textwidth]{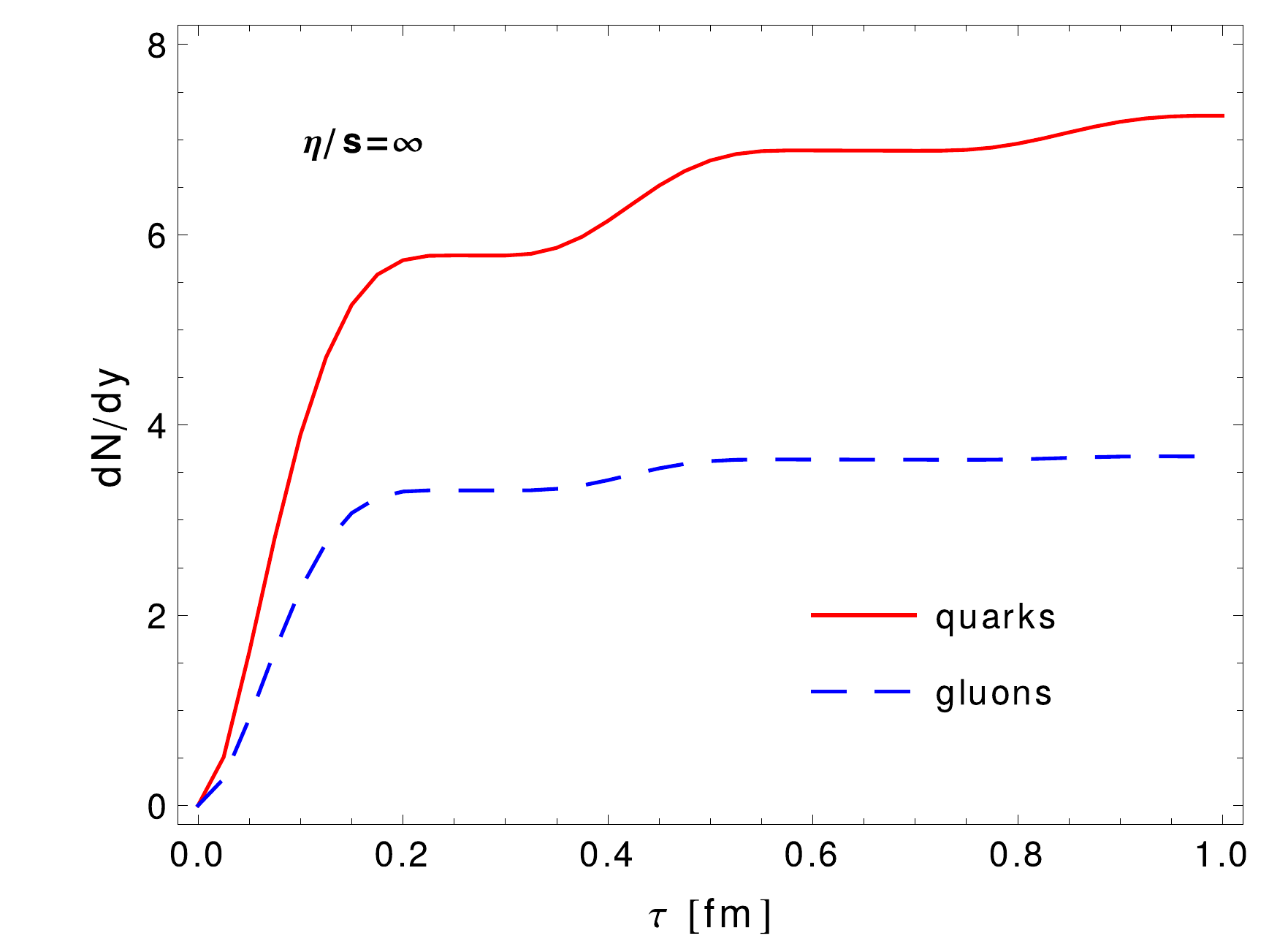}
\vspace{-0.2cm}
\end{center}
\caption{
Left panel: The transverse-momentum spectra of quarks (solid line) and gluons (dashed line) at $\tau=1$ fm. For comparison, the case of massive quarks is also presented (dotted line). Right panel: Proper-time dependence of the $\pt$-integrated rapidity density for quarks (solid line) and gluons (dashed line).
}
\label{qg}
\end{figure}
%
%
%
\begin{figure}[h]
\begin{center}
\includegraphics[width=0.49 \textwidth]{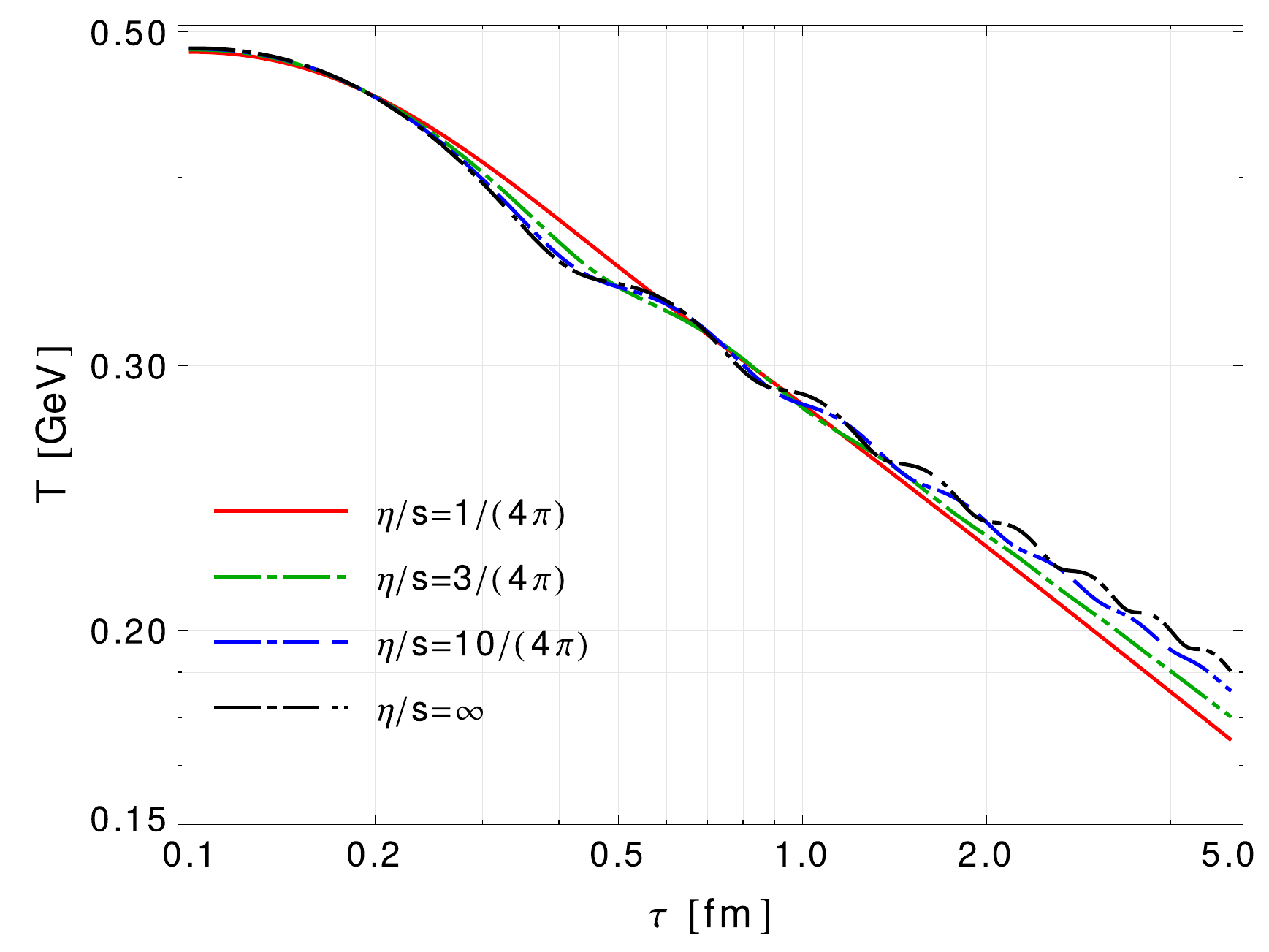}
 \includegraphics[width=0.49 \textwidth]{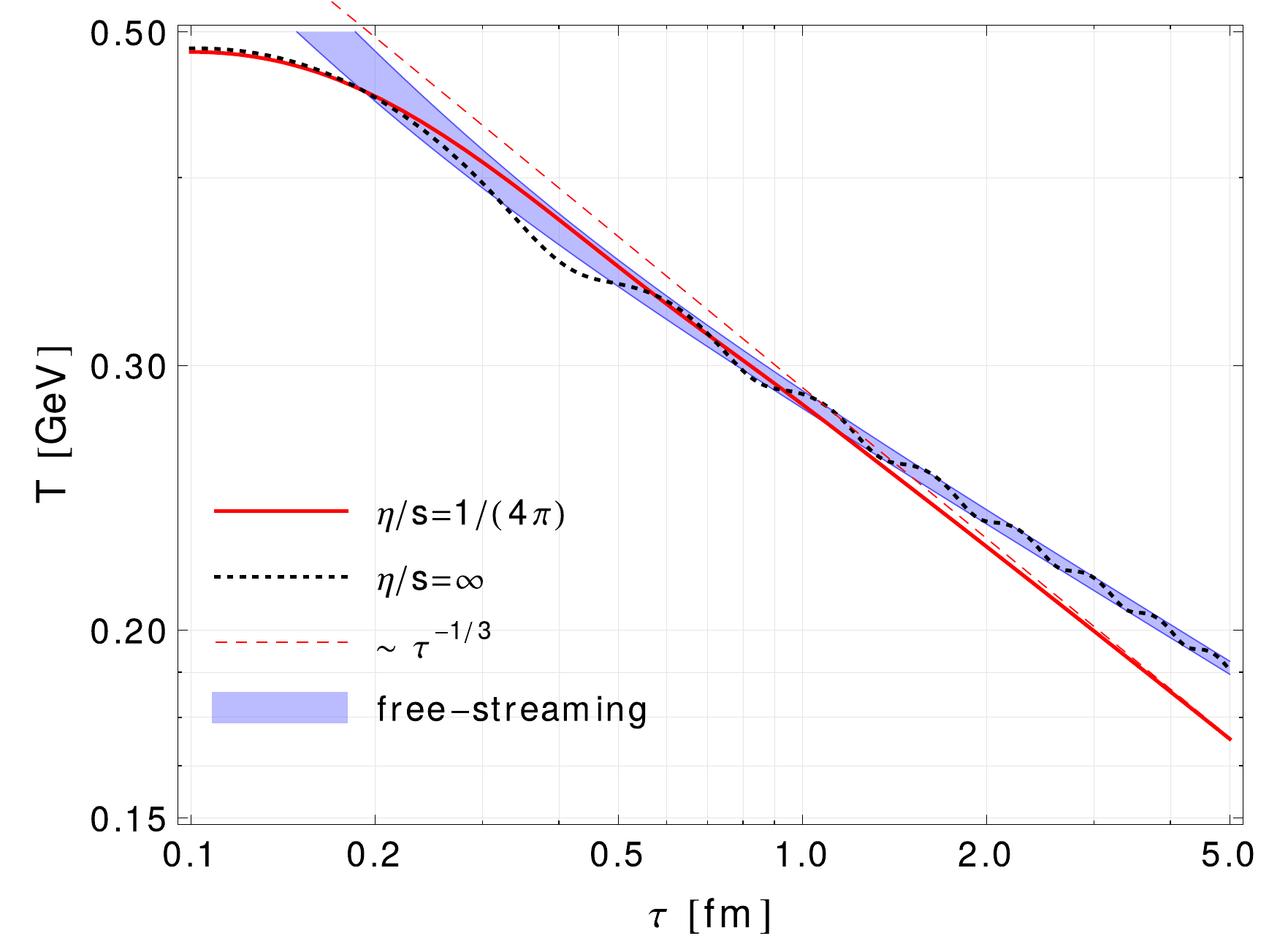}
\vspace{-0.2cm}
\end{center}
\caption{
Left panel: Proper-time dependence of the effective temperature in the system for various values of the shear viscosity to entropy density ratio. Right panel: The proper-time dependence of the effective temperature for $\eta/s = 1/(4\pi)$ (solid line) and $\eta/s = \infty$ (dotted line) compared to the Bjorken scaling (dashed line) and the free-streaming scaling (blue band). 
}
\label{T}
\end{figure}
%
%
\par We observe that after the evolution time \mbox{$\tau=0.5$ fm} the transverse-momentum spectrum is approximately exponential down to $\pt \approx 500$ MeV and exhibits a strong enhancement at low $\pt$. Exponential shape of the spectrum results from the tunneling in the changing/oscillatory color field \cite{Bialas:1999zg,Florkowski:2003mm}. It may be interpreted as  \textit{apparent thermalization of transverse degrees of freedom}\footnote{\textit{Apparent thermalization} in this case means that the exponential, thermal-like shape of the transverse-momentum spectrum results solely from the specific mechanism of particle production rather than from the particle collisions that gradually drive the system towards local equilibrium.}.  Nevertheless, one should keep in mind, that at the LHC energies considered herein (see also the results presented in Ref.~\cite{Ryblewski:2013eja} for $k=10$) the low-$\pt$ peak extends to higher values of the transverse momentum, as compared to the case of low energies considered in Ref.~\cite{Florkowski:2003mm}. In our case the  $\pt$ region below $500$ MeV consists of almost $45\%$ of the particles present in the system, while the particles with $\pt<1$ GeV  make up $\approx75\%$ of the total yield. Thus, the concept of apparent thermalization may be questionable in our case. 

It is interesting to note, that the low-$\pt$ peak may be traced back to the singularity in the quark production rate in Eq.~(\ref{rateq}) at $\pt=0$, that may be reduced if a finite, constituent quark mass is considered, see dotted line in the left panel of Fig.~\ref{qg} \footnote{The low-$\pt$ enhancement is not observed in Ref.~\cite{Ruggieri:2015yea}, where exclusively the gluon spectrum is considered.}. In the right panel of Fig.~\ref{qg} we present the proper-time dependence of the $\pt$-integrated parton yield per unit rapidity, separately for quarks (solid line) and gluons (dashed line). We see that quarks are more abundant than gluons, and both are produced mainly during the first fermi of the time evolution. As the color field decays below the parton production threshold, the particle production is strongly suppressed. After that time, partons just stream freely in the transverse direction, which is visible in the frozen transverse-momentum spectrum,  and oscillate longitudinally in the remnant color field (see the dotted line in the right panel of Fig.~1 from Ref.~\cite{Ryblewski:2013eja}).

\par Finally, we stress that the slope parameter $\lambda$ ($\approx 400$~MeV) read off from the right panel of Fig.~\ref{sp0} at $\tau=1$ fm  is quite different from the effective temperature $T$ ($\approx 300$~MeV)  calculated at the same proper time. The latter can be read off from the left panel of Fig.~\ref{T}. This difference may be easily understood in the following way. If $\eta/s$ is sufficiently large, the partons, once produced, are effectively free-streaming, see Eqs.~(\ref{kineq})-(\ref{kineg}). In this case, for the boost-invariant and transversally-homogeneous system,  the momentum-space anisotropy in the local rest frame, $\xi(\tau)$, evolves according to the formula $\xi(\tau) = (1+\xi(\tau_0))\left(\tau/\tau_0\right)^2 -1$ and the effective temperature, by the force of so called \textit{Landau matching} is $T(\tau) = R (\xi(\tau))^{1/4} \Lambda(\tau)$, where $\Lambda$ is the transverse temperature \cite{Martinez:2010sc,Florkowski:2010cf}. The latter is equivalent to  the slope parameter $\lambda$ (which in the case $\bar{\eta} = \infty$  is approximately constant for \mbox{$\tau > 1$ fm}), and the initial anisotropy parameter, $\xi(\tau_0)$, may be deduced from the relation $R_L\left(\xi(\tau_0)\right)/R_T\left(\xi(\tau_0)\right) = P_L(\tau_0)/P_T(\tau_0)$, where we choose $\tau_0 \ge 1$~fm. The functions $ R,  R_L,  R_T$ are defined in Ref.~\cite{Martinez:2010sc}, while longitudinal, $P_L(\tau)$, and transverse, $P_T(\tau)$, pressures are taken from Ref.~\cite{Ryblewski:2013eja}. The resulting  $T(\tau)$, including uncertainty connected with the choice of $\tau_0$, agrees very well with the case  $\eta/s=\infty$ in~Fig.~\ref{T}.
%
%
\begin{figure}[h]
\begin{center}
\includegraphics[width=0.49 \textwidth]{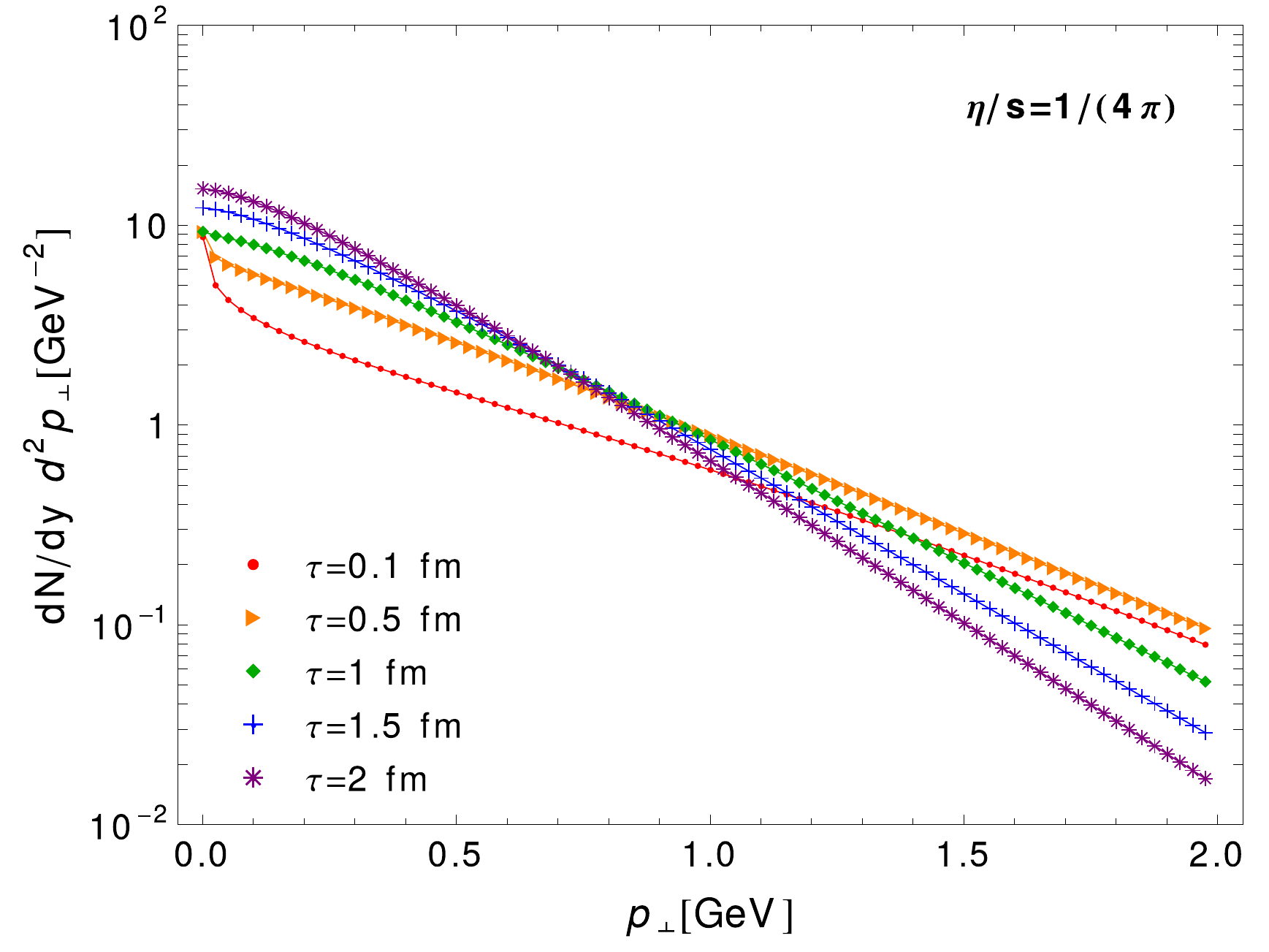} \includegraphics[width=0.49 \textwidth]{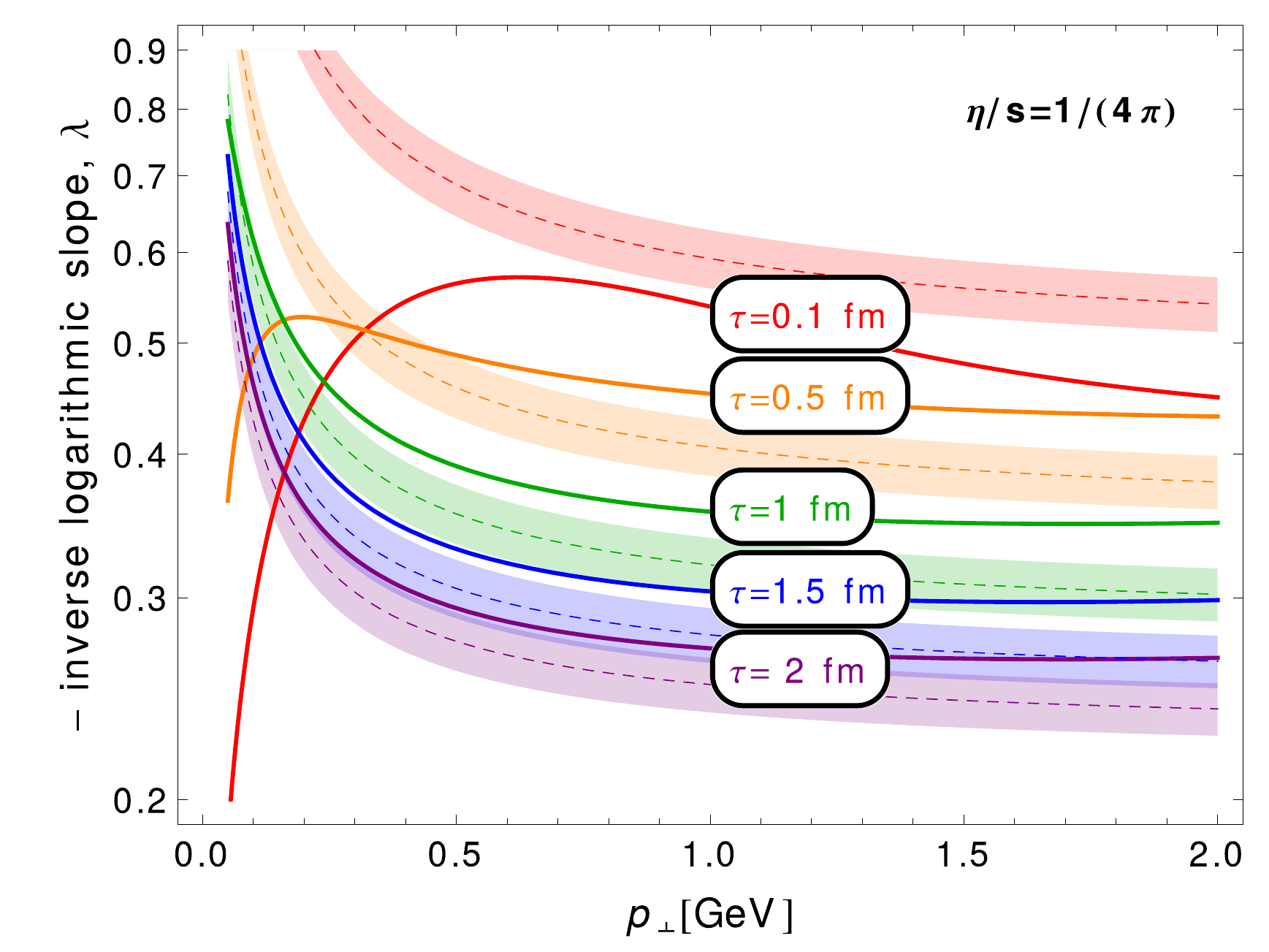}\\

\includegraphics[width=0.49 \textwidth]{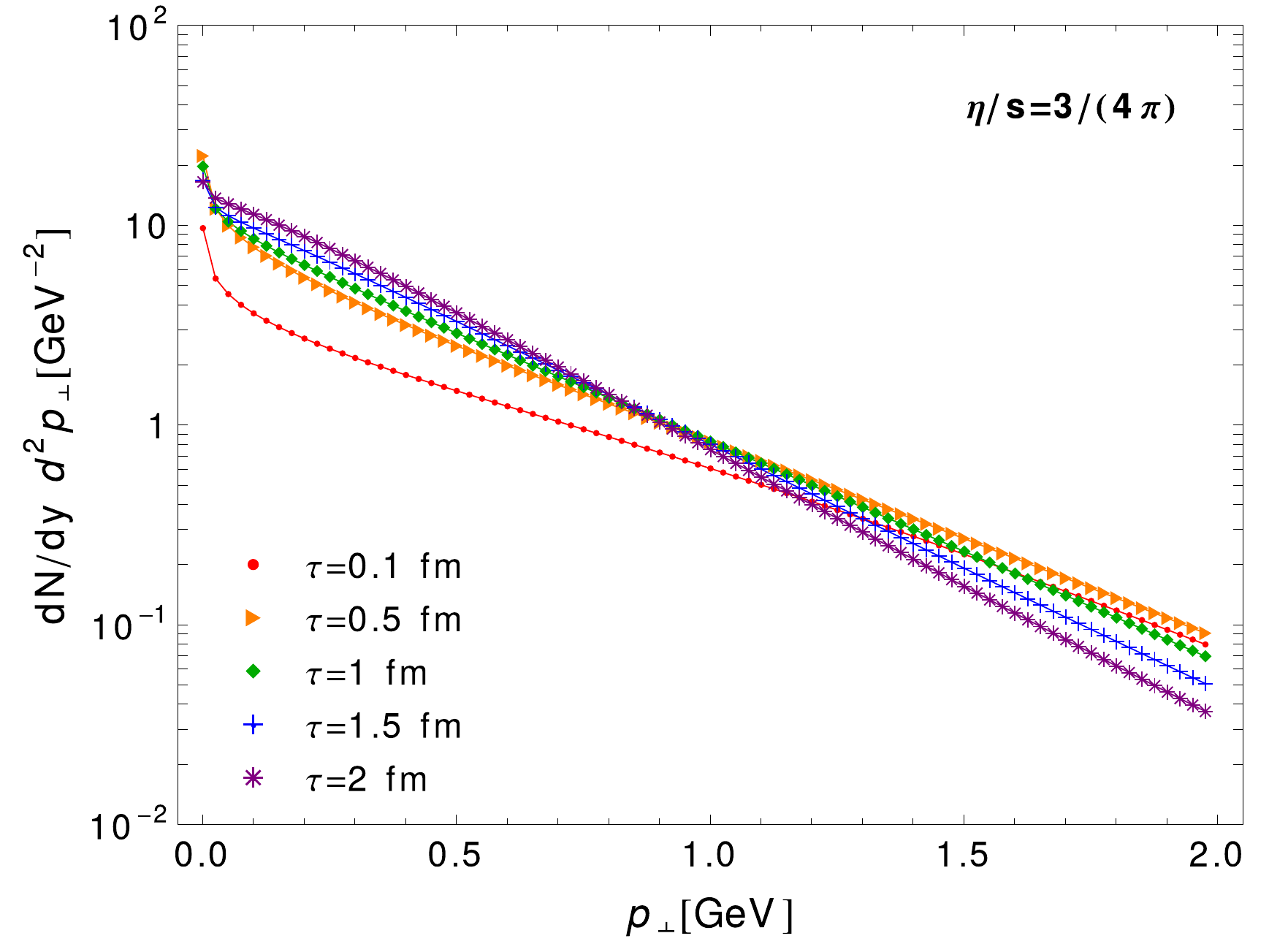}
\includegraphics[width=0.49 \textwidth]{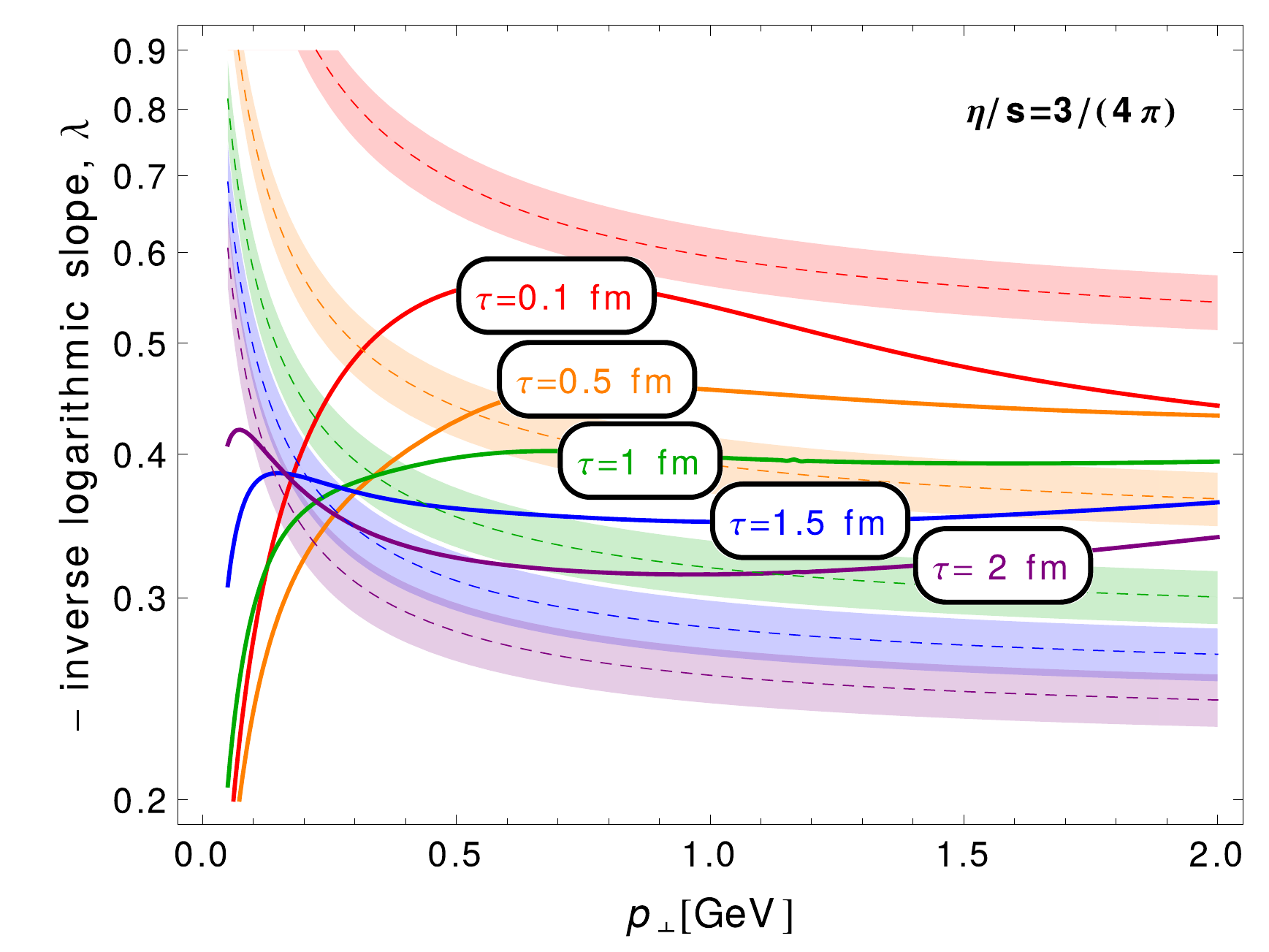}\\

\includegraphics[width=0.49 \textwidth]{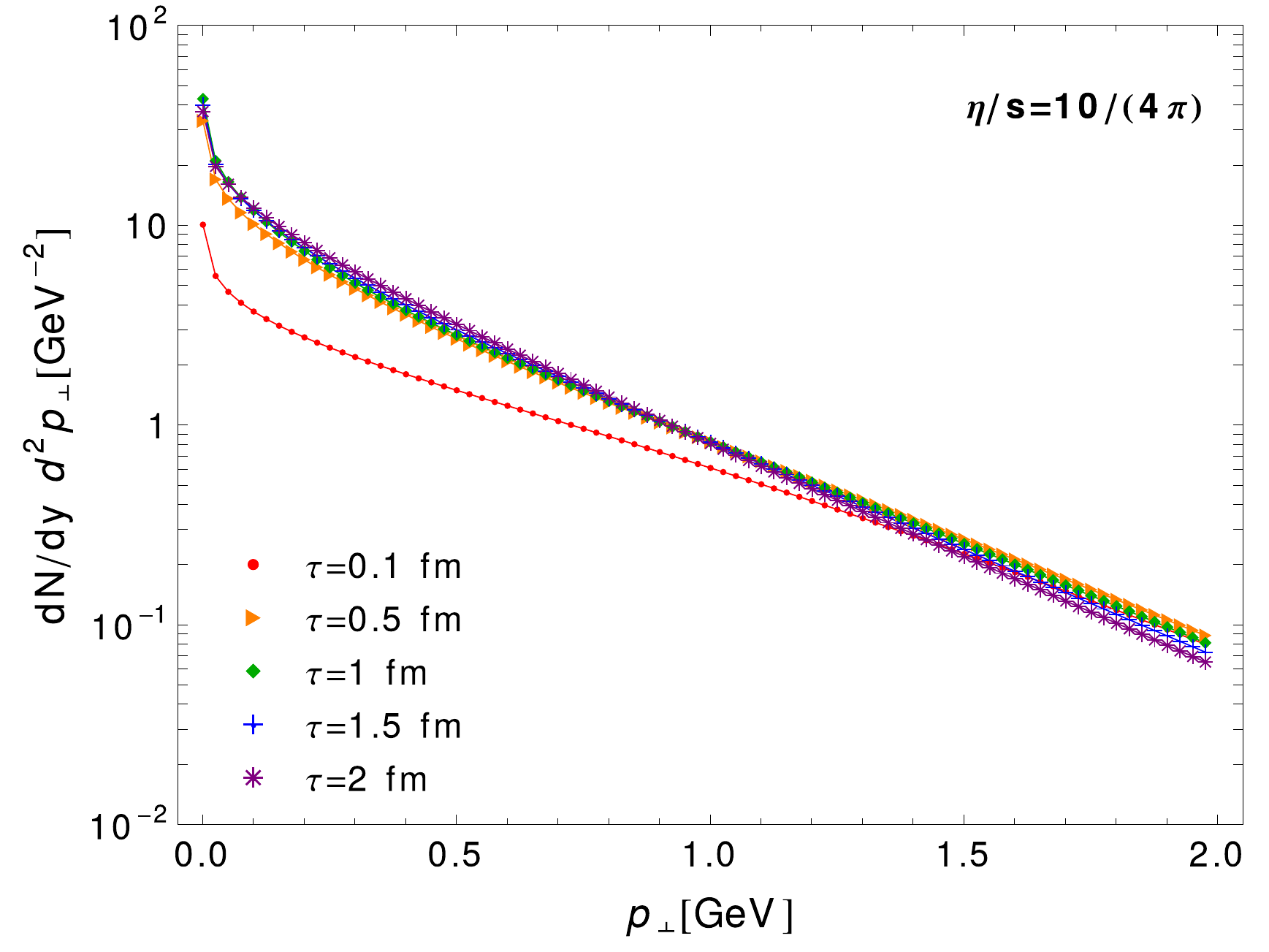}
\includegraphics[width=0.49 \textwidth]{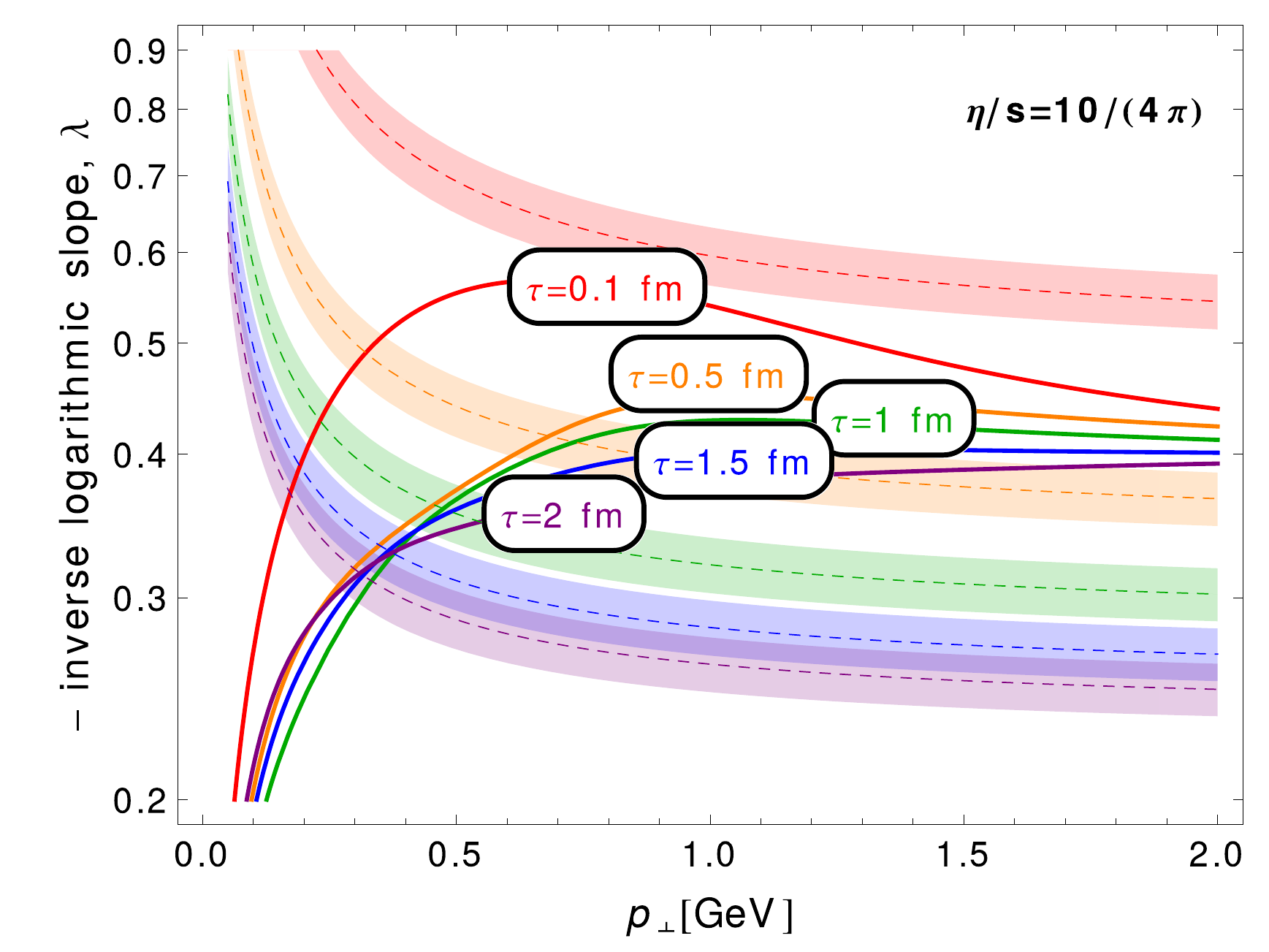}\\
\vspace{-0.2cm}
\end{center}
\caption{
Left panels: The $\pt$ spectra of partons calculated using Eq.~(\ref{cfmodel}) for $\eta/s = 1/4\pi$ (top), $\eta/s = 3/4\pi$ (middle), and $\eta/s = 10$ (bottom), and various freeze-out proper times. Right panels: Corresponding inverse logarithmic slopes of the spectra. The dashed lines in the top figure denote the slope of the fitted Boltzmann equilibrium spectrum.
}
\label{sp1}
\end{figure}
%
%
\subsection{Plasma with collisions}
\label{sect-col}
%
In this section we present the study of the parton production, including the collisions in the system, as calculated using Eq.~(\ref{cfmodel}). In the left panels of Fig.~\ref{sp1} we show the $\pt$ spectra of partons for a broad range of the shear viscosity to entropy density ratio, from $\eta/s = 1/(4\pi)$ (top), through $\eta/s = 3/(4\pi)$ (middle), to $\eta/s = 10/(4\pi)$ (bottom), and for various times of the proper-time at freeze-out (as indicated in the figures). We observe that the spectra have exponential, thermal-like shapes for all proper times and values of $\eta/s$ except for very early proper times, $\tau < 0.5$ fm, and/or the low momentum part of the spectra, $\pt < 500$~MeV. 

The exponential shape of the transverse-momentum spectra does not necessarily mean that the system is in local equilibrium. To conclude about the local equilibrium we have to analyse also the longitudinal spectra.  Alternatively, we may compare the local slope of the transverse-momentum spectrum with the local slope obtained for the three-dimensional equilibrated system characterised by the effective temperature obtained from the Landau matching. In view of the discussion in Section \ref{sect-colles}, in the right panels  of Fig.~\ref{sp1} we compare the local slopes of the $p_\perp$-spectra from the left panels (solid lines) with the local slopes resulting from Eq.~(\ref{cfeq}), i.e.,  for a locally equilibrated system of classical particles, $\lambda_{\rm eq} = T K_1 (\pt/T)/ K_0 (\pt/T)$ (dashed lines in the bands), where the effective $T(\tau)$ is presented in the left panel of Fig.~\ref{T} .   

For the lower bound $\eta/s = 1/(4\pi)$, see the right top panel of Fig.~\ref{sp1}, we observe that only for $\tau > 1$~fm the effective temperature of the system obtained from the Landau matching condition ($T=285$ MeV) gives the slope parameter $\lambda_{\rm eq}(\pt)$ well describing the actual $\lambda(\pt)$ (in order to judge the deviation of $\lambda_{\rm eq}$ from the exact $\lambda$ in Fig.~\ref{sp1} we introduced the error bands showing the change of $T$ by $\pm5$\%). This suggests that the plasma approaches full local equilibrium at the proper time of about 2 fm. On the other hand, at earlier times the spectra are off equilibrium, with the largest differences observed at low $\pt$. Thus, as expected, the inclusion of collisions specified by the lower bound of $\eta/s = 1/(4\pi)$ results in the almost complete thermalization of the system within $2$~fm. Moreover, the collisions, as included within RTA, seem to be more efficient in thermalizing  the system at low $\pt$, while, surprisingly,  large-$\pt$ part of the spectrum goes slightly off equilibrium once the collisions are included.

In the middle panels of Fig.~\ref{sp1} we present the analogous analysis as in top panels, however, here we consider $\eta/s = 3/(4\pi)$. We observe that the large value of the shear viscosity prevents the system from fast thermalization. It is especially visible at low momenta where the slope of the spectra deviates from the thermal one, compare the top panels. From the left panel of Fig.~\ref{T} we see that the effective temperature at $\tau=1$~fm is similar for all values of $\eta/s$. However, the slope in Fig.~\ref{sp1} deviates from this value significantly. We expect, however, that the observed anisotropies are small enough to be addressed within the perturbative viscous fluid dynamical modeling. 

Finally, in the bottom panels of Fig.~\ref{sp1} we present the case of $\eta/s = 10/(4\pi)$. In this case the system does not thermalize, which may be again deduced from the slope, which does not decrease in time fast enough, as compared to the dashed lines shown in the figure. Altogether, this case is similar to the case described in Section \ref{sect-colles}. Large deviations from equilibrium suggest that in this case in order to describe the system within an effective fluid dynamical picture it may be necessary to use more sophisticated approaches, for instance anisotropic hydrodynamics \cite{Martinez:2010sc,Florkowski:2010cf}.
%
\subsection{Thermalization versus hydrodynamization}
\label{sect-col}
%
\par The last point which we want to address is the question whether, although not completely thermalized, the system presented above may be still reasonably well described within some effective  dissipative fluid dynamical framework, for instance with the standard viscous fluid dynamics by including perturbative corrections to the local equilibrium distribution given by Eq.~(\ref{feq}).
The latter concept is referred to as the \textit{hydrodynamization} of the system and it was first proposed by Heller et al. in Ref.~\cite{Heller:2011ju}.
\par In the  (0+1)D case the distribution function ansatz for conformal (massless) classical system of particles within viscous fluid dynamics has the form \cite{Strickland:2014pga}
\begin{eqnarray}
f^{\rm visc}(x,p) &=& f^{\rm eq}(x,p)\left[1 + \frac{p_\mu \pi ^{\mu\nu} p_\nu}{2(\varepsilon+P)T^2}\right] \nonumber\\
  &=& f^{\rm eq}\left(\frac{E}{T}\right)\left[1 + \frac{3 \pi}{16\, \varepsilon \,T^2}\left(\pt^2 -2 \pl^2\right)\right],
\label{fvisc}
\end{eqnarray}
where $\pi^{\mu\nu} ={\rm diag}\left(0, \pi/2, \pi/2, -\pi\right)$, $\pi = 2(P_\perp - P_\parallel)/3$, and $\varepsilon$ and  $P$ are equilibrium energy density and pressure, respectively. Inserting formula (\ref{fvisc}) into Eq.~(\ref{cfbi}) we may subsequently  obtain respective $\lambda_{\rm visc}$ using Eq.~(\ref{slope}),
\begin{eqnarray}
\lambda_{\rm visc}  &=& T \frac{((a \hat{p}_\perp)^{-1}+ \hat{p}_\perp) K_1(\hat{p}_\perp) -2 K_2(\hat{p}_\perp)}{((a \hat{p}_\perp)^{-1}+ \hat{p}_\perp) K_0(\hat{p}_\perp) -4 K_1(\hat{p}_\perp)},
\label{lvisc}
\end{eqnarray}
with $\hat{p}_\perp=\pt/T$ and $a = (P_T - P_L)/(8\,\varepsilon)$. In the left panel of Fig.~\ref{sp1} the results of Eq.~(\ref{lvisc}) (thick dotted lines) are compared with the values of $\lambda$ for the case $\eta/s=1/(4\pi)$ (solid lines) and with the equilibrium results $\lambda_{\rm eq}$ (dashed lines). We clearly see that the inclusion of viscous corrections improves significantly the description of the transient non-equilibrium behaviour of the exact results at large $\pt$. Finally, in the right panel of Fig.~\ref{sp1} we show that the values of $\lambda_{\rm visc}$ are close to those of $\lambda$ also in the case of $\eta/s=3/(4\pi)$ (compare the respective case in Fig.~\ref{sp1}).
%
%
\begin{figure}[t]
\begin{center}
\includegraphics[width=0.49 \textwidth]{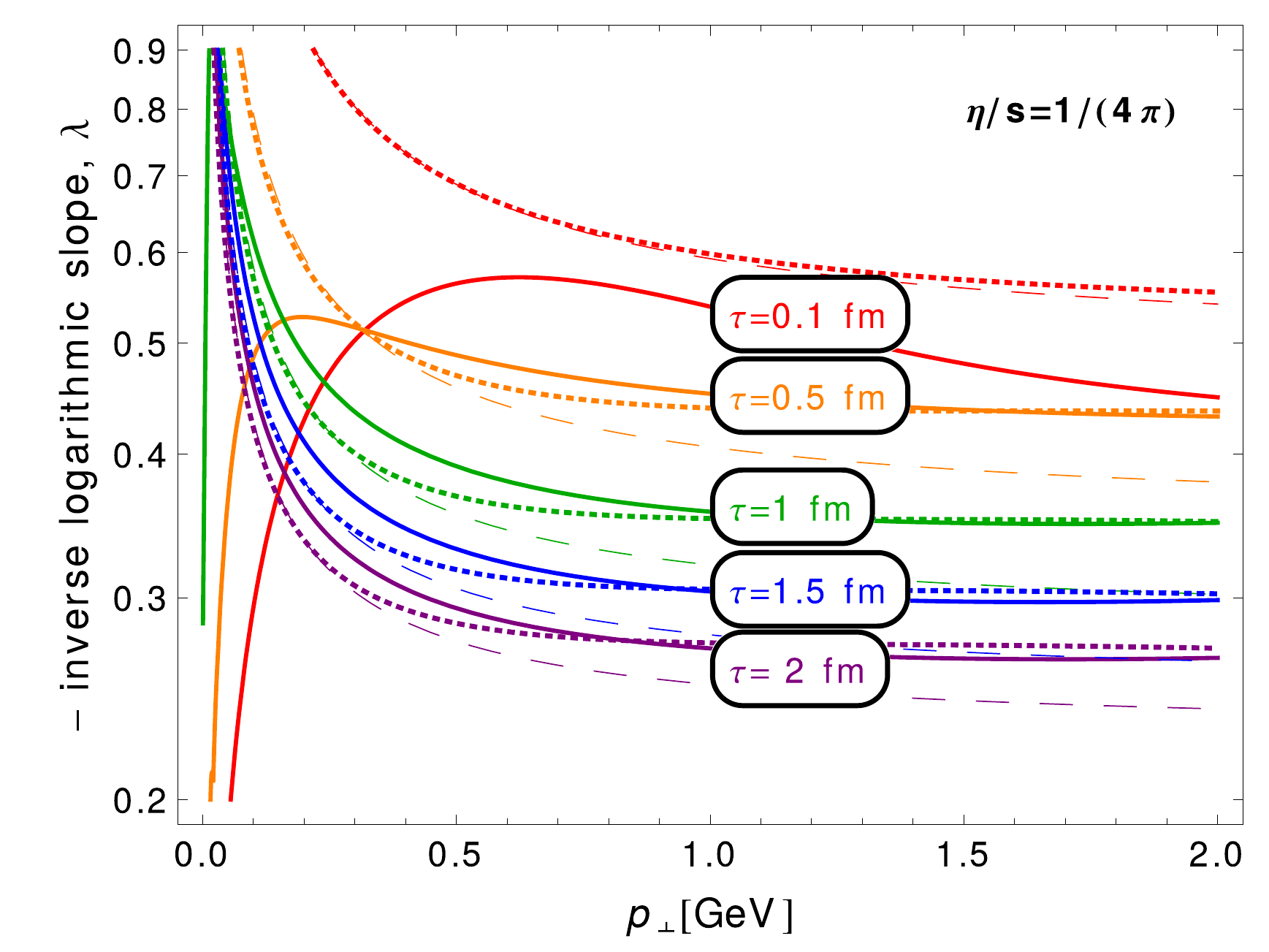} 
\includegraphics[width=0.49 \textwidth]{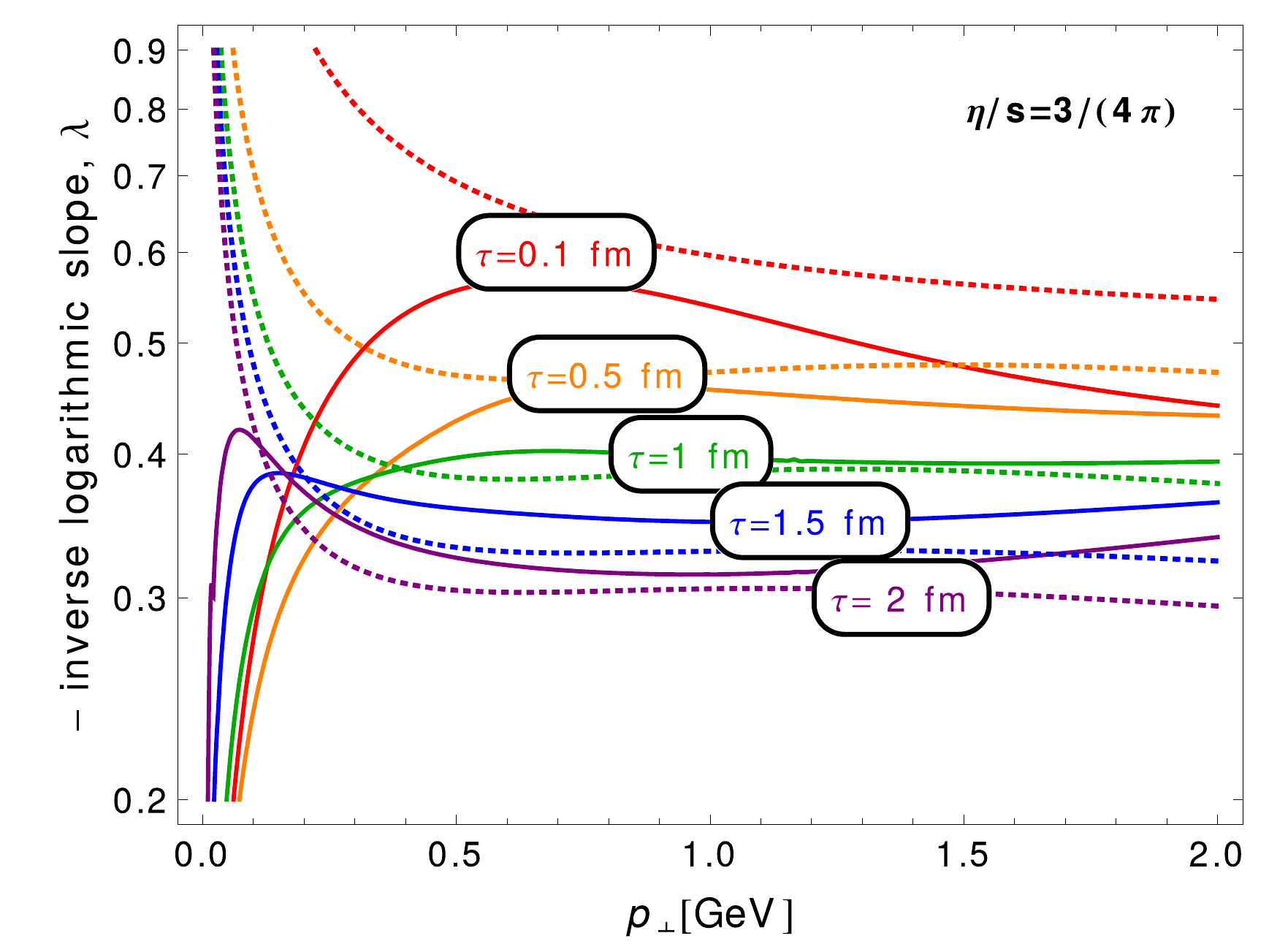}
\vspace{-0.2cm}
\end{center}
\caption{
Left panel: The slope parameters $\lambda$ as a function of $\pt$ (solid lines) at various proper times for the case $\eta/s =1/(4\pi)$ compared with the results obtained within viscous fluid dynamic slope $\lambda_{\rm visc}$ (dotted lines). To guide the reader's eye we show also the results for equilibrium case $\lambda_{\rm eq}$. Right panel: Same as left panel except for the case $\eta/s =3/(4\pi)$. For the sake of clarity the results for $\lambda_{\rm eq}$ are removed here.
}
\label{sp2}
\end{figure}
%
%
\section{Conclusions}
\label{sect-con}
%
In this paper, extending  the scope of Ref.~\cite{Ryblewski:2013eja}, we have presented a detailed study of the proper-time dependence of the spectra of partons produced in the color-flux-tube model. The calculations have been performed for different ratios of the shear viscosity to the entropy density. We have studied the interplay between the production of particles by the Shwinger tunneling process and their equilibration due to the collisions. We have found that the collisions in the system are required to thermalize the system eventually in all three directions in momentum space. Otherwise, the parton spectrum, although apparently thermal in the transverse direction,  does not correspond to a true local equilibrium state.  In the collisionless case the effective temperature of the system is shown to drop accordingly to the free-streaming solution found within anisotropic hydrodynamics framework. If the collisions are included, the value $\eta/s \approx 1/(4\pi)$ is necessary to bring the system to the full local equilibrium state within $\approx 2$ fm. On the other hand, the hydrodynamization of the system, that is the time when the fluid dynamical framework is applicable, is achieved within less than $1$ fm. 

%
\section*{Acknowledgments}
%
                           
The author would like to thank Wojciech Florkowski and Leonardo Tinti for interesting discussions during the preparation of this work and the critical reading of the manuscript. The research was supported by the National Science Center  Grant  No. DEC-2012/07/D/ST2/02125 and the Polish Ministry of Science and Higher Education fellowship for young scientists (X edition).

\bigskip



\bibliographystyle{iopart-num}
\bibliography{refs}

\providecommand{\newblock}{}
\begin{thebibliography}{100}
\expandafter\ifx\csname url\endcsname\relax
  \def\url#1{{\tt #1}}\fi
\expandafter\ifx\csname urlprefix\endcsname\relax\def\urlprefix{URL }\fi
\providecommand{\eprint}[2][]{\url{#2}}

\bibitem{Shuryak:1978ij}
Shuryak E~V 1978 {\em Phys. Lett.\/} {\bf B78} 150 [Yad. Fiz.28,796(1978)]

\bibitem{Muronga:2003ta}
Muronga A 2004 {\em Phys. Rev.\/} {\bf C69} 034903 (\textit{Preprint}
  \eprint{nucl-th/0309055})

\bibitem{Heinz:2005bw}
Heinz U~W, Song H and Chaudhuri A~K 2006 {\em Phys. Rev.\/} {\bf C73} 034904
  (\textit{Preprint} \eprint{nucl-th/0510014})

\bibitem{Bhalerao:2005mm}
Bhalerao R~S, Blaizot J~P, Borghini N and Ollitrault J~Y 2005 {\em Phys.
  Lett.\/} {\bf B627} 49--54 (\textit{Preprint} \eprint{nucl-th/0508009})

\bibitem{Baier:2006um}
Baier R, Romatschke P and Wiedemann U~A 2006 {\em Phys. Rev.\/} {\bf C73}
  064903 (\textit{Preprint} \eprint{hep-ph/0602249})

\bibitem{Baier:2007ix}
Baier R, Romatschke P, Son D~T, Starinets A~O and Stephanov M~A 2008 {\em
  JHEP\/} {\bf 04} 100 (\textit{Preprint} \eprint{0712.2451})

\bibitem{Dusling:2007gi}
Dusling K and Teaney D 2008 {\em Phys. Rev.\/} {\bf C77} 034905
  (\textit{Preprint} \eprint{0710.5932})

\bibitem{Broniowski:2008vp}
Broniowski W, Chojnacki M, Florkowski W and Kisiel A 2008 {\em Phys. Rev.
  Lett.\/} {\bf 101} 022301 (\textit{Preprint} \eprint{0801.4361})

\bibitem{NoronhaHostler:2008ju}
Noronha-Hostler J, Noronha J and Greiner C 2009 {\em Phys. Rev. Lett.\/} {\bf
  103} 172302 (\textit{Preprint} \eprint{0811.1571})

\bibitem{Bozek:2009dw}
Bozek P 2010 {\em Phys. Rev.\/} {\bf C81} 034909 (\textit{Preprint}
  \eprint{0911.2397})

\bibitem{El:2009vj}
El A, Xu Z and Greiner C 2010 {\em Phys. Rev.\/} {\bf C81} 041901
  (\textit{Preprint} \eprint{0907.4500})

\bibitem{Alver:2010dn}
Alver B~H, Gombeaud C, Luzum M and Ollitrault J~Y 2010 {\em Phys. Rev.\/} {\bf
  C82} 034913 (\textit{Preprint} \eprint{1007.5469})

\bibitem{Martinez:2010sc}
Martinez M and Strickland M 2010 {\em Nucl. Phys.\/} {\bf A848} 183--197
  (\textit{Preprint} \eprint{1007.0889})

\bibitem{Florkowski:2010cf}
Florkowski W and Ryblewski R 2011 {\em Phys. Rev.\/} {\bf C83} 034907
  (\textit{Preprint} \eprint{1007.0130})

\bibitem{PeraltaRamos:2010je}
Peralta-Ramos J and Calzetta E 2010 {\em Phys. Rev.\/} {\bf C82} 054905
  (\textit{Preprint} \eprint{1003.1091})

\bibitem{Petersen:2010cw}
Petersen H, Qin G~Y, Bass S~A and Muller B 2010 {\em Phys. Rev.\/} {\bf C82}
  041901 (\textit{Preprint} \eprint{1008.0625})

\bibitem{Denicol:2010tr}
Denicol G~S, Kodama T and Koide T 2010 {\em J. Phys.\/} {\bf G37} 094040
  (\textit{Preprint} \eprint{1002.2394})

\bibitem{Pratt:2010jt}
Pratt S and Torrieri G 2010 {\em Phys. Rev.\/} {\bf C82} 044901
  (\textit{Preprint} \eprint{1003.0413})

\bibitem{Schenke:2010rr}
Schenke B, Jeon S and Gale C 2011 {\em Phys. Rev. Lett.\/} {\bf 106} 042301
  (\textit{Preprint} \eprint{1009.3244})

\bibitem{Niemi:2011ix}
Niemi H, Denicol G~S, Huovinen P, Molnar E and Rischke D~H 2011 {\em Phys. Rev.
  Lett.\/} {\bf 106} 212302 (\textit{Preprint} \eprint{1101.2442})

\bibitem{Shen:2011eg}
Shen C, Heinz U, Huovinen P and Song H 2011 {\em Phys. Rev.\/} {\bf C84} 044903
  (\textit{Preprint} \eprint{1105.3226})

\bibitem{Ryblewski:2012rr}
Ryblewski R and Florkowski W 2012 {\em Phys. Rev.\/} {\bf C85} 064901
  (\textit{Preprint} \eprint{1204.2624})

\bibitem{Pang:2012he}
Pang L, Wang Q and Wang X~N 2012 {\em Phys. Rev.\/} {\bf C86} 024911
  (\textit{Preprint} \eprint{1205.5019})

\bibitem{Karpenko:2015xea}
Karpenko I~A, Huovinen P, Petersen H and Bleicher M 2015 {\em Phys. Rev.\/}
  {\bf C91} 064901 (\textit{Preprint} \eprint{1502.01978})

\bibitem{Bhalerao:2015iya}
Bhalerao R~S, Jaiswal A and Pal S 2015 {\em Phys. Rev.\/} {\bf C92} 014903
  (\textit{Preprint} \eprint{1503.03862})

\bibitem{Becattini:2015ska}
Becattini F, Inghirami G, Rolando V, Beraudo A, Del~Zanna L, De~Pace A, Nardi
  M, Pagliara G and Chandra V 2015 {\em Eur. Phys. J.\/} {\bf C75} 406
  (\textit{Preprint} \eprint{1501.04468})

\bibitem{Bazow:2015cha}
Bazow D, Heinz U~W and Martinez M 2015 {\em Phys. Rev.\/} {\bf C91} 064903
  (\textit{Preprint} \eprint{1503.07443})

\bibitem{Tinti:2015xwa}
Tinti L 2015  (\textit{Preprint} \eprint{1506.07164})

\bibitem{McLerran:1993ni}
McLerran L~D and Venugopalan R 1994 {\em Phys. Rev.\/} {\bf D49} 2233--2241
  (\textit{Preprint} \eprint{hep-ph/9309289})

\bibitem{Iancu:2003xm}
Iancu E and Venugopalan R 2003 {The Color glass condensate and high-energy
  scattering in QCD} {\em {In *Hwa, R.C. (ed.) et al.: Quark gluon plasma*
  249-3363}\/} (\textit{Preprint} \eprint{hep-ph/0303204})

\bibitem{JalilianMarian:2005jf}
Jalilian-Marian J and Kovchegov Y~V 2006 {\em Prog. Part. Nucl. Phys.\/} {\bf
  56} 104--231 (\textit{Preprint} \eprint{hep-ph/0505052})

\bibitem{Gelis:2009wh}
Gelis F, Lappi T and McLerran L 2009 {\em Nucl. Phys.\/} {\bf A828} 149--160
  (\textit{Preprint} \eprint{0905.3234})

\bibitem{Rybczynski:2013yba}
Rybczynski M, Stefanek G, Broniowski W and Bozek P 2014 {\em Comput. Phys.
  Commun.\/} {\bf 185} 1759--1772 (\textit{Preprint} \eprint{1310.5475})

\bibitem{Schenke:2012wb}
Schenke B, Tribedy P and Venugopalan R 2012 {\em Phys. Rev. Lett.\/} {\bf 108}
  252301 (\textit{Preprint} \eprint{1202.6646})

\bibitem{Mrowczynski:1993qm}
Mrowczynski S 1993 {\em Phys. Lett.\/} {\bf B314} 118--121

\bibitem{Randrup:2003cw}
Randrup J and Mrowczynski S 2003 {\em Phys. Rev.\/} {\bf C68} 034909
  (\textit{Preprint} \eprint{nucl-th/0303021})

\bibitem{Romatschke:2003ms}
Romatschke P and Strickland M 2003 {\em Phys. Rev.\/} {\bf D68} 036004
  (\textit{Preprint} \eprint{hep-ph/0304092})

\bibitem{Arnold:2003rq}
Arnold P~B, Lenaghan J and Moore G~D 2003 {\em JHEP\/} {\bf 08} 002
  (\textit{Preprint} \eprint{hep-ph/0307325})

\bibitem{Mrowczynski:2004kv}
Mrowczynski S, Rebhan A and Strickland M 2004 {\em Phys. Rev.\/} {\bf D70}
  025004 (\textit{Preprint} \eprint{hep-ph/0403256})

\bibitem{Rebhan:2008uj}
Rebhan A, Strickland M and Attems M 2008 {\em Phys. Rev.\/} {\bf D78} 045023
  (\textit{Preprint} \eprint{0802.1714})

\bibitem{Ipp:2010uy}
Ipp A, Rebhan A and Strickland M 2011 {\em Phys. Rev.\/} {\bf D84} 056003
  (\textit{Preprint} \eprint{1012.0298})

\bibitem{Gelis:2013rba}
Epelbaum T and Gelis F 2013 {\em Phys. Rev. Lett.\/} {\bf 111} 232301
  (\textit{Preprint} \eprint{1307.2214})

\bibitem{Berges:2013eia}
Berges J, Boguslavski K, Schlichting S and Venugopalan R 2014 {\em Phys.
  Rev.\/} {\bf D89} 074011 (\textit{Preprint} \eprint{1303.5650})

\bibitem{Chesler:2010bi}
Chesler P~M and Yaffe L~G 2011 {\em Phys. Rev. Lett.\/} {\bf 106} 021601
  (\textit{Preprint} \eprint{1011.3562})

\bibitem{CaronHuot:2011dr}
Caron-Huot S, Chesler P~M and Teaney D 2011 {\em Phys. Rev.\/} {\bf D84} 026012
  (\textit{Preprint} \eprint{1102.1073})

\bibitem{Heller:2011ju}
Heller M~P, Janik R~A and Witaszczyk P 2012 {\em Phys. Rev. Lett.\/} {\bf 108}
  201602 (\textit{Preprint} \eprint{1103.3452})

\bibitem{vanderSchee:2013pia}
van~der Schee W, Romatschke P and Pratt S 2013 {\em Phys. Rev. Lett.\/} {\bf
  111} 222302 (\textit{Preprint} \eprint{1307.2539})

\bibitem{Ryblewski:2013eja}
Ryblewski R and Florkowski W 2013 {\em Phys. Rev.\/} {\bf D88} 034028
  (\textit{Preprint} \eprint{1307.0356})

\bibitem{Casher:1978wy}
Casher A, Neuberger H and Nussinov S 1979 {\em Phys. Rev.\/} {\bf D20} 179--188

\bibitem{Glendenning:1983qq}
Glendenning N~K and Matsui T 1983 {\em Phys. Rev.\/} {\bf D28} 2890--2891

\bibitem{Bialas:1984wv}
Bialas A and Czyz W 1984 {\em Phys. Rev.\/} {\bf D30} 2371

\bibitem{Bialas:1984ap}
Bialas A and Czyz W 1985 {\em Z. Phys.\/} {\bf C28} 255

\bibitem{Bialas:1985is}
Bialas A and Czyz W 1986 {\em Nucl. Phys.\/} {\bf B267} 242

\bibitem{Gyulassy:1986jq}
Gyulassy M and Iwazaki A 1985 {\em Phys. Lett.\/} {\bf B165} 157--161

\bibitem{Gatoff:1987uf}
Gatoff G, Kerman A~K and Matsui T 1987 {\em Phys. Rev.\/} {\bf D36} 114

\bibitem{Heinz:1983nx}
Heinz U~W 1983 {\em Phys. Rev. Lett.\/} {\bf 51} 351

\bibitem{Heinz:1984yq}
Heinz U~W 1985 {\em Annals Phys.\/} {\bf 161} 48

\bibitem{Elze:1986qd}
Elze H~T, Gyulassy M and Vasak D 1986 {\em Nucl. Phys.\/} {\bf B276} 706--728

\bibitem{Elze:1986hq}
Elze H~T, Gyulassy M and Vasak D 1986 {\em Phys. Lett.\/} {\bf B177} 402--408

\bibitem{Bialas:1986mt}
Bialas A and Czyz W 1986 {\em Acta Phys. Polon.\/} {\bf B17} 635

\bibitem{Bialas:1987en}
Bialas A, Czyz W, Dyrek A and Florkowski W 1988 {\em Nucl. Phys.\/} {\bf B296}
  611

\bibitem{Dyrek:1988eb}
Dyrek A and Florkowski W 1989 {\em Nuovo Cim.\/} {\bf A102} 1013

\bibitem{Ruggieri:2015yea}
Ruggieri M, Puglisi A, Oliva L, Plumari S, Scardina F and Greco V 2015
  (\textit{Preprint} \eprint{1505.08081})

\bibitem{Plumari:2012ep}
Plumari S, Puglisi A, Scardina F and Greco V 2012 {\em Phys. Rev.\/} {\bf C86}
  054902 (\textit{Preprint} \eprint{1208.0481})

\bibitem{Scardina:2014gxa}
Scardina F, Perricone D, Plumari S, Ruggieri M and Greco V 2014 {\em Phys.
  Rev.\/} {\bf C90} 054904 (\textit{Preprint} \eprint{1408.1313})

\bibitem{Scardina:2014nja}
Scardina F, Ruggieri M, Plumari S and Greco V 2014 {\em Nucl. Phys.\/} {\bf
  A932} 484--489

\bibitem{Puglisi:2014sha}
Puglisi A, Plumari S and Greco V 2014 {\em Phys. Rev.\/} {\bf D90} 114009
  (\textit{Preprint} \eprint{1408.7043})

\bibitem{Das:2015aga}
Das S~K, Ruggieri M, Mazumder S, Greco V and Alam J~e 2015 {\em J. Phys.\/}
  {\bf G42} 095108 (\textit{Preprint} \eprint{1501.07521})

\bibitem{Ruggieri:2015tsa}
Ruggieri M, Plumari S, Scardina F and Greco V 2015 {\em Nucl. Phys.\/} {\bf
  A941} 201--211 (\textit{Preprint} \eprint{1502.04596})

\bibitem{Plumari:2015cfa}
Plumari S, Guardo G~L, Scardina F and Greco V 2015 {\em Phys. Rev.\/} {\bf C92}
  054902 (\textit{Preprint} \eprint{1507.05540})

\bibitem{Schwinger:1951nm}
Schwinger J~S 1951 {\em Phys. Rev.\/} {\bf 82} 664--679

\bibitem{Brezin:1970xf}
Brezin E and Itzykson C 1970 {\em Phys. Rev.\/} {\bf D2} 1191--1199

\bibitem{Schmidt:1998vi}
Schmidt S~M, Blaschke D, Ropke G, Smolyansky S~A, Prozorkevich A~V and Toneev
  V~D 1998 {\em Int. J. Mod. Phys.\/} {\bf E7} 709--722 (\textit{Preprint}
  \eprint{hep-ph/9809227})

\bibitem{Fukushima:2009er}
Fukushima K, Gelis F and Lappi T 2009 {\em Nucl. Phys.\/} {\bf A831} 184--214
  (\textit{Preprint} \eprint{0907.4793})

\bibitem{Gelis:2013oca}
Gelis F and Tanji N 2013 {\em Phys. Rev.\/} {\bf D87} 125035 (\textit{Preprint}
  \eprint{1303.4633})

\bibitem{Gelis:2015kya}
Gelis F and Tanji N 2015  (\textit{Preprint} \eprint{1510.05451})

\bibitem{Bhatnagar:1954zz}
Bhatnagar P~L, Gross E~P and Krook M 1954 {\em Phys. Rev.\/} {\bf 94} 511--525

\bibitem{Baym:1984np}
Baym G 1984 {\em Phys. Lett.\/} {\bf B138} 18--22

\bibitem{Baym:1985tna}
Baym G 1984 {\em Nucl. Phys.\/} {\bf A418} 525C--537C

\bibitem{Banerjee:1989by}
Banerjee B, Bhalerao R~S and Ravishankar V 1989 {\em Phys. Lett.\/} {\bf B224}
  16

\bibitem{Heiselberg:1995sh}
Heiselberg H and Wang X~N 1996 {\em Phys. Rev.\/} {\bf C53} 1892--1902
  (\textit{Preprint} \eprint{hep-ph/9504244})

\bibitem{Wong:1996va}
Wong S~M~H 1996 {\em Phys. Rev.\/} {\bf C54} 2588--2599 (\textit{Preprint}
  \eprint{hep-ph/9609287})

\bibitem{Nayak:1996ex}
Nayak G~C and Ravishankar V 1997 {\em Phys. Rev.\/} {\bf D55} 6877--6886
  (\textit{Preprint} \eprint{hep-th/9610215})

\bibitem{Nayak:1997kp}
Nayak G~C and Ravishankar V 1998 {\em Phys. Rev.\/} {\bf C58} 356--364
  (\textit{Preprint} \eprint{hep-ph/9710406})

\bibitem{Bhalerao:1999hj}
Bhalerao R~S and Nayak G~C 2000 {\em Phys. Rev.\/} {\bf C61} 054907
  (\textit{Preprint} \eprint{hep-ph/9907322})

\bibitem{Policastro:2001yc}
Policastro G, Son D~T and Starinets A~O 2001 {\em Phys. Rev. Lett.\/} {\bf 87}
  081601 (\textit{Preprint} \eprint{hep-th/0104066})

\bibitem{Kovtun:2004de}
Kovtun P, Son D~T and Starinets A~O 2005 {\em Phys. Rev. Lett.\/} {\bf 94}
  111601 (\textit{Preprint} \eprint{hep-th/0405231})

\bibitem{Bjorken:1982qr}
Bjorken J~D 1983 {\em Phys. Rev.\/} {\bf D27} 140--151

\bibitem{Huang:1982ik}
Huang K 1982 {\em {QUARKS, LEPTONS AND GAUGE FIELDS}\/}

\bibitem{Anderson1974466}
Anderson J and Witting H 1974 {\em Physica\/} {\bf 74} 466 -- 488 ISSN
  0031-8914
  \urlprefix\url{http://www.sciencedirect.com/science/article/pii/0031891474903553}

\bibitem{Czyz:1986mr}
Czyz W and Florkowski W 1986 {\em Acta Phys. Polon.\/} {\bf B17} 819--837

\bibitem{Dyrek:1986vv}
Dyrek A and Florkowski W 1987 {\em Phys. Rev.\/} {\bf D36} 2172

\bibitem{Romatschke:2011qp}
Romatschke P 2012 {\em Phys. Rev.\/} {\bf D85} 065012 (\textit{Preprint}
  \eprint{1108.5561})

\bibitem{Florkowski:2013lza}
Florkowski W, Ryblewski R and Strickland M 2013 {\em Nucl. Phys.\/} {\bf A916}
  249--259 (\textit{Preprint} \eprint{1304.0665})

\bibitem{Florkowski:2013lya}
Florkowski W, Ryblewski R and Strickland M 2013 {\em Phys. Rev.\/} {\bf C88}
  024903 (\textit{Preprint} \eprint{1305.7234})

\bibitem{Cooper:1974mv}
Cooper F and Frye G 1974 {\em Phys. Rev.\/} {\bf D10} 186

\bibitem{Misner:1974qy}
Misner C~W, Thorne K~S and Wheeler J~A 1973 {\em {Gravitation}\/} (San
  Francisco: W. H. Freeman)

\bibitem{Bialas:1999zg}
Bialas A 1999 {\em Phys. Lett.\/} {\bf B466} 301--304 (\textit{Preprint}
  \eprint{hep-ph/9909417})

\bibitem{Florkowski:2003mm}
Florkowski W 2004 {\em Acta Phys. Polon.\/} {\bf B35} 799--808
  (\textit{Preprint} \eprint{nucl-th/0309049})

\bibitem{Strickland:2014pga}
Strickland M 2014 {\em Acta Phys. Polon.\/} {\bf B45} 2355--2394
  (\textit{Preprint} \eprint{1410.5786})

\end{thebibliography}


\end{document}